\documentclass[fleqn]{article}
\usepackage{graphicx} 
\usepackage{geometry}
\usepackage[T1]{fontenc}
\usepackage[latin9]{inputenc}
\usepackage{float}
\usepackage{xfrac}
\usepackage{amsmath}
\usepackage{amssymb}
\usepackage{amsfonts}
\usepackage{xcolor}
\usepackage{hyphenat}
\usepackage[backend=biber, citestyle=numeric-comp, sorting=none, url = false, isbn = false, eprint=false]{biblatex}
\usepackage{titlesec}
\usepackage{mathtools}
\usepackage{xr}
\usepackage{booktabs}
\usepackage{subcaption}
\usepackage{array,multirow,tabularx,lipsum}
\usepackage{booktabs}
\usepackage{setspace}
\usepackage{longtable}
\usepackage{cancel}
\usepackage[most]{tcolorbox}
\usepackage{scalerel}
\usepackage[normalem]{ulem}
\usepackage[hidelinks]{hyperref}
\usepackage{longtable}\usepackage{xparse}
\usepackage{authblk}

\NewDocumentCommand{\DIV}{om}{%
  \IfValueT{#1}{\setcounter{#2}{\numexpr#1-1\relax}}%
  \csname #2\endcsname
}

\newcommand{\stkout}[1]{\ifmmode\text{\sout{\ensuremath{#1}}}\else\sout{#1}\fi}

\makeatletter
\newcommand\subsubsubsection{\@startsection{paragraph}{4}{\z@}{-2.5ex\@plus -1ex \@minus -.25ex}{1.25ex \@plus .25ex}{\normalfont\large\bfseries}}
\newcommand\subsubsubsubsection{\@startsection{subparagraph}{5}{\z@}{-2.5ex\@plus -1ex \@minus -.25ex}{1.25ex \@plus .25ex}{\normalfont\normalsize\bfseries}}
\makeatother

\def\be{\begin{equation}}
\def\ee{\end{equation}}
\def\bea{\begin{eqnarray}}
\def\eea{\end{eqnarray}}

\def\be{\begin{equation}}
\def\ee{\end{equation}}
\def\bea{\begin{eqnarray}}
\def\eea{\end{eqnarray}}

\def\a{\alpha}
\def\b{\beta}

\def\x{{\bf x}}

\def\wo{\setminus}

\newcolumntype{P}[1]{>{\raggedright}p{#1}}

\usepackage{xcolor}

\begin{document}

\title{\textbf{
Resource supply dynamics control stability and chaos in complex ecosystems 
}}

\author[1]{Jamila Rowland-Chandler}
\author[2, *]{Akshit Goyal}
\author[1, *]{Wenying Shou}
\affil[1]{Centre for Life's Origins and Evolution, Department of Genetics, Evolution and Environment, University College London, WC1E 6BT, United Kingdom}
\affil[2]{International Centre for Theoretical Sciences, Tata Institute of Fundamental Research, Bengaluru 560089, India}
\affil[*]{Equal contribution corresponding author (order determined by a coin flip). 

Electronic addresses: W. Shou: \url{wenying.shou@gmail.com}; A. Goyal: \url{akshitg@icts.res.in}
}

\maketitle

\begin{abstract}

Ecological interactions are often mediated by feedbacks between organisms and their resource environments. 
Yet, how resource supply dynamics dictate collective dynamical phases of an ecosystem remains unclear. 
Here, we analyse a generalised consumer--resource model with non-reciprocal interactions to demonstrate that self-renewing versus externally-supplied resources yield fundamentally different dynamical phase diagrams. 
As interactions become increasingly non-reciprocal, ecosystems relying on self-renewing resources transition from stable dynamics to chaos and ultimately to infeasibility. 
By contrast, ecosystems with externally-supplied resources remain stable over a broader parameter range and transition to infeasibility without experiencing an intervening chaotic phase. 
Using the cavity method, we derive a unified stability condition applicable to a broad class of resource supply functions, explaining why externally-supplied resources can expand the stable region.
We show that stability hinges crucially on the susceptibility of resources to perturbations, which depends strongly on their supply.
Further, we show that external resource supply suppresses chaos in the unstable region by drastically reducing the susceptibility of resources closest to extinction.
Our findings demonstrate that resource dynamics fundamentally reshape the accessible dynamical behaviours of an ecosystem, with implications for interpreting microbial community experiments.

\end{abstract}

A central goal of ecology is to understand how collective dynamical behaviours in ecosystems, from stable equilibria to chaotic fluctuations ~\cite{faith_long-term_2013, lozupone_diversity_2012, fernandez_how_1999, fernandez-gonzalez_microbial_2016, hu_emergent_2022, ratzke_strength_2020, martin-platero_high_2018, beninca_chaos_2008, beninca_species_2015, ross_metabolic_2024}, emerge from interactions among many species.
Recent approaches from statistical physics have made significant progress on this problem by treating ecosystems as large disordered dynamical systems, revealing novel dynamical phases and phase transitions controlled by the statistical properties of ecological interactions ~\cite{allesina_predicting_2015, may_will_1972, may_qualitative_1973, bunin_ecological_2017, mallmin_chaotic_2024, hu_emergent_2022, grilli_modularity_2016, blumenthal_phase_2024, liu_complex_2025, liu_ecosystem_2024, arnoulx_de_pirey_many-species_2024, dalmedigos_dynamical_2020, roy_complex_2020, arnoulx_de_pirey_critical_2025, allesina_stability_2012, altieri_properties_2021, mahadevan_spatiotemporal_2023, rowland-chandler_resource_2025}.

Much of this work has focused on generalised Lotka--Volterra (gLV) models with direct pairwise species interactions ~\cite{allesina_stability_2012, may_will_1972, may_qualitative_1973, bunin_ecological_2017, mallmin_chaotic_2024, hu_emergent_2022, grilli_modularity_2016, arnoulx_de_pirey_many-species_2024, roy_complex_2020, arnoulx_de_pirey_critical_2025, altieri_properties_2021, mahadevan_spatiotemporal_2023}, which lack explicit feedbacks between organisms and their environments that often mediate species interactions ~\cite{niehaus_microbial_2019, culp_cross-feeding_2023, dal_bello_resourcediversity_2021, estrela_nutrient_2021, flint_interactions_2007, goldford_emergent_2018, hoek_resource_2016, lawrence_species_2012, ritchie_predictions_1993, ross_metabolic_2024, sun_metabolic_2024, tylianakis_resource_2008, wong_fluid_2023, lee_resource_2023, daniels_changes_2023,momeni_lotka-volterra_2017, shou_synthetic_2007, violle_phylogenetic_2011}.
This limitation has motivated growing interest in consumer--resource models (CRM) where species interactions emerge from competition and cooperation for shared resources ~\cite{blumenthal_phase_2024, liu_complex_2025, liu_ecosystem_2024, dalmedigos_dynamical_2020, macarthur_competition_1964, macarthur_species_1969, chesson_macarthurs_1990, advani_statistical_2018, goldford_emergent_2018, dandrea_emergent_2020, gibbs_stability_2022, cui_effect_2020, cui_diverse_2021, marsland_available_2019, marsland_minimal_2020, rowland-chandler_resource_2025}.
Recent work has shown that decreasing the reciprocity between resource depletion and consumer growth can drive a transition from stability to chaotic abundance fluctuations~\cite{blumenthal_phase_2024, liu_complex_2025, liu_ecosystem_2024,rowland-chandler_resource_2025}. 
However, these models assume self-renewing resources; whether these results extend to externally supplied resources remains unclear, despite some studies claim that they do ~\cite{liu_complex_2025, liu_ecosystem_2024}.
As a result, we lack a general understanding of how resource supply modes shape the dynamical phases of ecosystems, hindering interpretation of recent experiments linking resource supplies to dynamical fluctuations ~\cite{hu_emergent_2022, gore_transition_2025, ratzke_strength_2020}.

In this Letter, we show that changing resource supply mode fundamentally alters the phase diagram of dynamical states in a complex consumer--resource ecosystem.
As interaction reciprocity decreases, while ecosystems with self-renewing resources transition from stability to chaotic boom--bust dynamics and ultimately to infeasibility, ecosystems with externally-supplied resources remain stable over a wider parameter range and transition directly to infeasibility.
Using the cavity method, we derive a unified stability condition for a broad class of consumer--resource models, and show that different resource supply modes lead to distinct stability conditions depending on whether resources can go extinct.
Our results demonstrate that the mode of resource supply can determine which dynamical behaviours an ecosystem can exhibit.

\begin{figure*}[t!]
\includegraphics[width=0.9\textwidth]{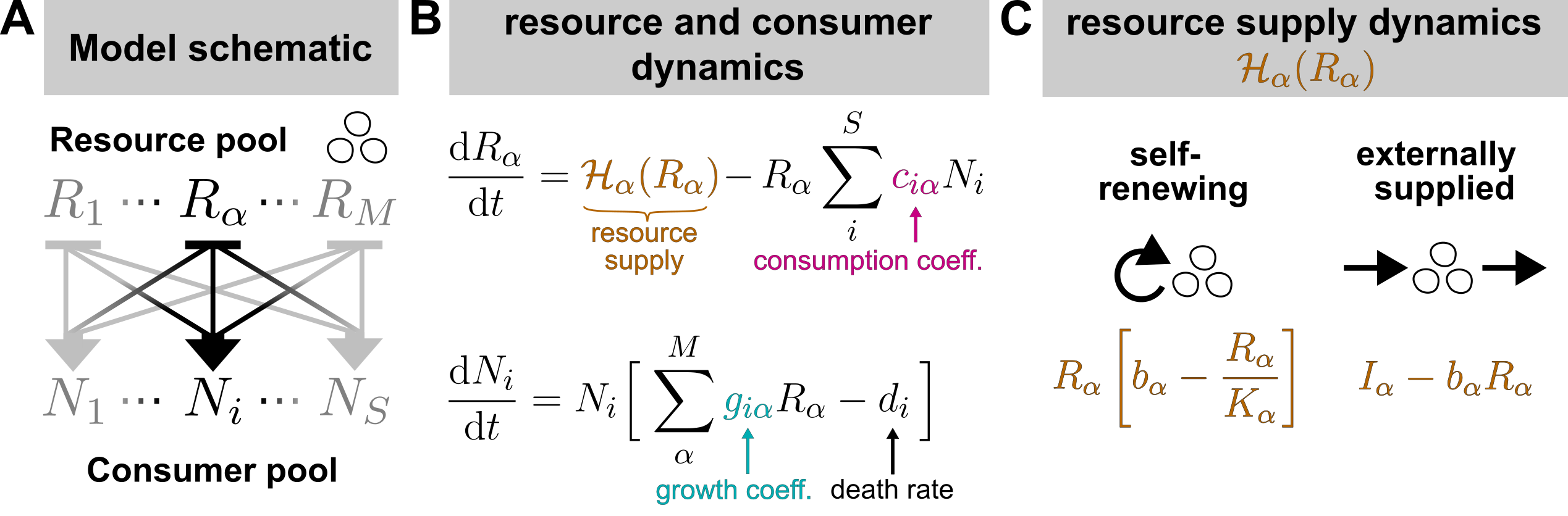}
\caption{
\scriptsize{\textsf{\textbf{Schematic for generalised consumer-resource models with non-reciprocal interactions.}
\textbf{(A)} Interactions between the resource and consumer pool.
\textbf{(B)} General form of resource and consumer dynamics.
\textbf{(C)} Resource supply functions investigated in this paper.
}}}
\label{fig:1}
\end{figure*}

\vskip 5pt
\subsection*{Model}We consider a generalised consumer-resource model with non-reciprocal interactions, where a pool of $S$ consumer species grow from consuming $M$ substitutable resources (Fig.~\ref{fig:1}).
The dynamics of resource $\alpha$ ($R_\alpha$, $\alpha \in \{1, ... M\}$) and consumer $i$ ($N_i$, $i \in \{1, ... S\}$) are described by the following set of differential equations:

\begin{eqnarray}
    \frac{\mathrm{d} R_\alpha}{\mathrm{d} t} &&= \mathcal{H}_\alpha (R_\alpha) - R_\alpha\sum_i^S c_{i \alpha} N_i,
    \label{eq:res_dynamics}
    \\
    \frac{\mathrm{d} N_i}{\mathrm{d} t} &&= N_i \left[ \sum_\alpha^M g_{i \alpha} R_\alpha - d_i\right].
    \label{eq:sp_dynamics}
\end{eqnarray}

Here, $c_{i \alpha}$ is the per-capita consumption rate of resource $\alpha$ by consumer $i$, $g_{i \alpha}$ is the corresponding growth rate of consumer $i$ on resource $\alpha$, and $d_i$ is consumer $i$'s death rate.
$\mathcal{H}_\alpha (R_\alpha)$ is the resource supply function, which describes resource dynamics in the absence of consumers.
When resources are self-renewing or biological (e.g., bacteria are resources for phage), $\mathcal{H}_\alpha (R_\alpha)$ describes logistic growth:

\begin{equation}
    \text{self-renewing:}\quad\mathcal{H}_\alpha (R_\alpha) = R_\alpha \left[ b_\alpha - \frac{R_\alpha}{K_\alpha} \right],
    \label{eq:self-renewing_supply}
\end{equation}

where $b_\alpha$ is the intrinsic growth rate of resource $\alpha$ and $K_\alpha$ is its carrying capacity.
When resources are externally-supplied or abiotic (e.g., chemicals supplied in a chemostat for bacterial growth), $\mathcal{H}_\alpha (R_\alpha)$ takes the form 

\begin{equation}
    \text{externally-supplied:}\quad\mathcal{H}_\alpha (R_\alpha) = I_\alpha - b_\alpha R_\alpha,
    \label{eq:externally-supplied_supply}
\end{equation}
where $I_\alpha$ is the influx rate of resource $\alpha$, independent of resource density, and $b_\alpha$ is now interpreted as the outflux or dilution rate.
When resource dynamics are much faster than consumer dynamics, both classes of dynamics reduce to gLV model and can become indistinguishable under certain conditions (SI Appendix E). 
Here, we investigate the opposite limit where resource environments actively shape species interactions, and resource dynamics cannot be ignored.

\vskip 5pt
\subsection*{Parametrisation}To study the typical dynamics of highly-diverse communities ($M,S\gg 1$), following prior literature, we sample model parameters as random variables~\cite{advani_statistical_2018,may_will_1972,bunin_ecological_2017,allesina_stability_2012} drawn from fixed distributions:\label{parametrisation} 

\begin{equation}
\begin{aligned}
I_\alpha &= \mu_I+\sigma_I z_{I,\alpha}, \qquad
b_\alpha = \mu_b+\sigma_b z_{b,\alpha}, \\
K_\alpha &= \mu_K+\sigma_K z_{K,\alpha},\\
c_{i\alpha} &= \frac{\mu_c}{M}+\frac{\sigma_c}{\sqrt{M}} \left[\rho z_{g,i\alpha}+\sqrt{1-\rho^2}\,z_{c,i\alpha}\right],\\
g_{i\alpha} &= \frac{\mu_g}{M}+\frac{\sigma_g}{\sqrt{M}} z_{g,i\alpha},\\
d_i &= \mu_d+\sigma_d z_{d,i}, \qquad
\gamma^{-1}=\frac{S}{M}.
\end{aligned}
\end{equation}
where all $z_{\langle\cdot\rangle}$ terms represent uncorrelated standard Gaussian random variables (note that our results also generalise to non-Gaussian distributions; SI Appendix B).
Consumption $(c_{i \alpha})$ and growth $(g_{i \alpha})$ rates are scaled by the resource pool size $M$ to ensure a proper thermodynamic limit ($M\to\infty$, $S\to\infty$) while maintaining a finite ratio $\gamma^{-1}$.
For simplicity, we set $\mu_c = \mu_g \equiv \mu$ and $\sigma_c = \sigma_g \equiv \sigma$ (quantifying heterogeneity) in all simulations.
The parameter $\rho={\rm corr}(c_{i\alpha},g_{i\alpha})$ represents the reciprocity of consumer--resource interactions: when $\rho = 1$, $c_{i \alpha} = g_{i \alpha}$, whereas when $\rho = 0$ they are independent.

\vskip 5pt
\subsection*{\label{results_simulation} Resource supply mode sets accessible dynamical phases}
\begin{figure*}[t!]
\includegraphics[width=0.9\textwidth]{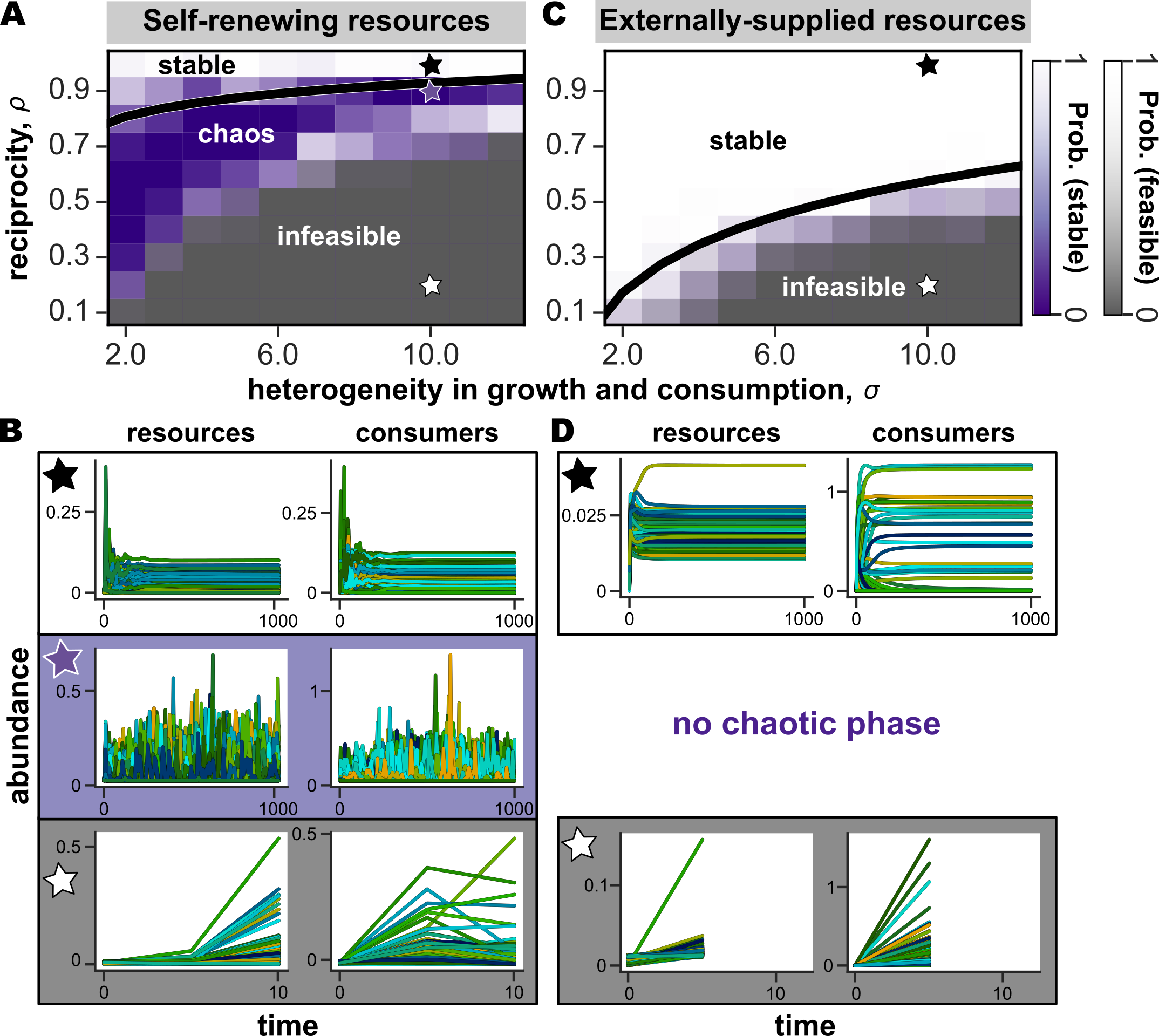}
\caption{
\scriptsize{\textsf{\textbf{The mode of resource supply governs accessible dynamical phases and the size of stable region.}
\textbf{(A, B)} Consumer-resource model with self-renewing resources. (A) Phase diagram. 
Each cell shows results from 20 simulated communities, with darker purple indicating a higher proportion of simulations showing unstable but feasible dynamics (maximum Lyapunov exponent $> 0$), darker grey indicating a higher proportion showing infeasible dynamics (simulations are numerically unstable and thus terminate early), and white indicating stable communities.
Black line = analytically-derived stability condition $\rho^2 = S^*/M^*$, where $S^*/M^*$ was estimated from simulations.
(B) Representative dynamics from the stable (black star, top), chaotic (purple star, middle) and infeasible regions (white star, bottom).
Each curve in each plot represents a single consumer or resource.
\textbf{(C)} Phase diagram for the model with externally-supplied resources.
Black line = analytically-derived stability condition in a simple limit, $\rho^2 = S^*/M$, where $S^*/M$ was estimated from simulations.
\textbf{(D)} Representative dynamics from the stable and infeasible regions of the externally-supplied model.
There is no chaotic phase.
(Simulation parameters are given in SI Appendix D.)}}}
\label{fig:2}
\end{figure*}

Previous work on consumer-resource models with self-renewing resources showed that the dynamical phase diagram can be plotted in  terms of the reciprocity $\rho$ and heterogeneity $\sigma$ in consumption and growth~\cite{blumenthal_phase_2024,liu_complex_2025,rowland-chandler_resource_2025}.
Decreasing $\rho$ (or increasing $\sigma$) drives communities through two successive transitions: from stable to chaotic dynamics, and then to infeasible dynamics where abundances could grow unbounded (Fig.~\ref{fig:2}A, B);~\cite{blumenthal_phase_2024}). 
To test if changing the mode of resource supply impacts stability
we simulated both self-renewing and externally-supplied resource models and plotted the dynamical phase diagram, taking care to match consumer and resource statistics in both cases (SI Appendix D).

Surprisingly, the externally-supplied model remained stable over a far greater range of $\rho$ and $\sigma$ than the self-renewing model (Fig.~\ref{fig:2}C).
Moreover, when the externally-supplied model did lose stability, it bypassed the chaotic regime entirely and instead transitioned from globally stable directly to infeasible dynamics (Fig \ref{fig:2}C, D).
Any chaotic trajectories observed near the stability-infeasibility boundary (Fig. \ref{fig:2}C, light purple) were likely due to finite size effects, and needed parameters to be extremely fine-tuned to be observed
(SI Appendix E Fig. S2).
These results demonstrate that ecosystems with externally-supplied resources have qualitatively different collective dynamics than with self-renewing resources.

\begin{figure*}[t!]
\includegraphics[width=0.9\textwidth]{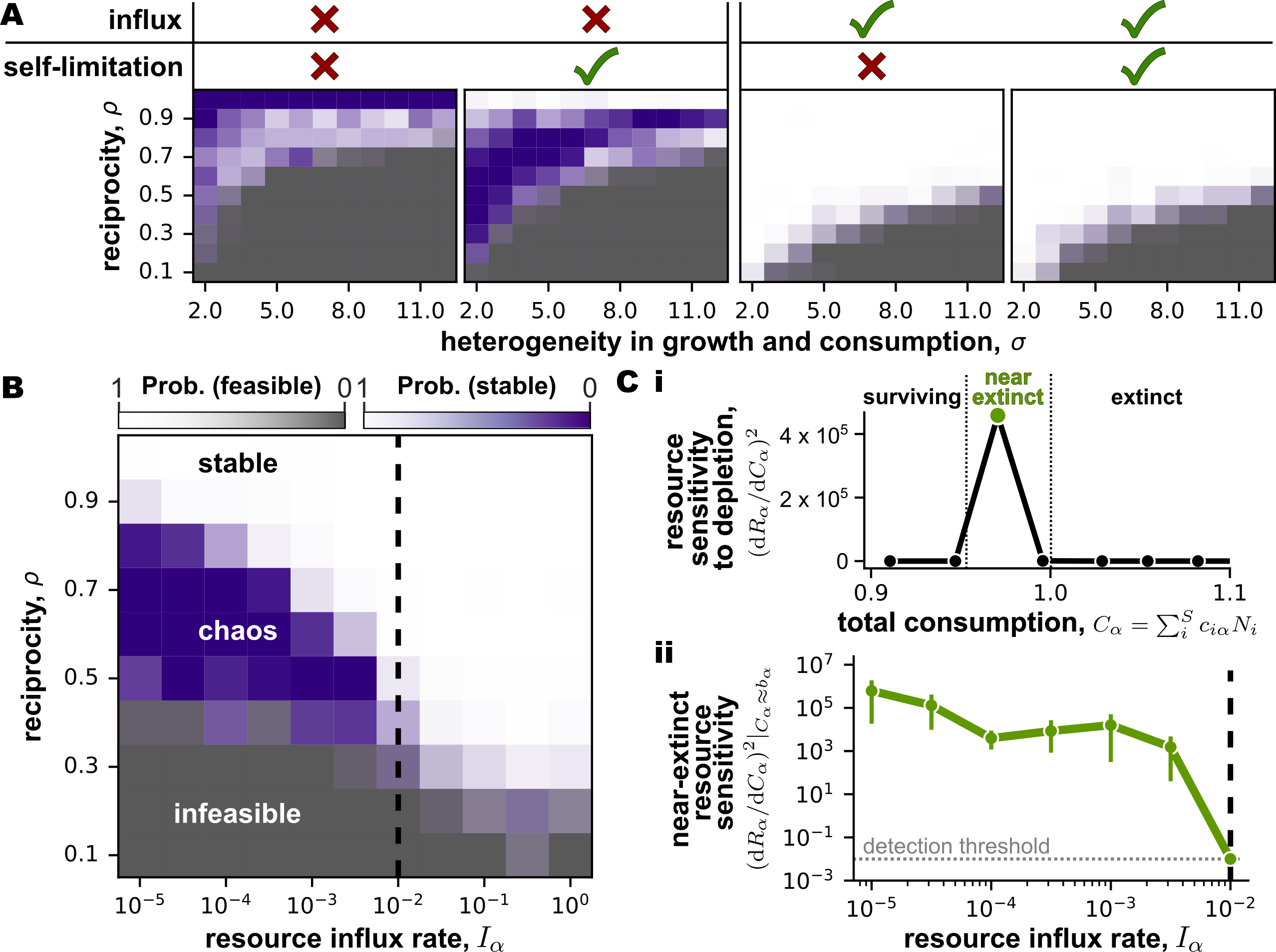}
\caption{
\scriptsize{\textsf{\textbf{External influx of resources suppresses chaos.}
(Simulation parameters are given in SI Appendix D.)
\textbf{(A)} Phase diagrams of the hybrid model, where resources are both externally-supplied and self-renewing: $\mathcal{H}_\alpha (R_\alpha) = I_\alpha + R_\alpha \left[ b_\alpha - R_\alpha / K_\alpha \right]$.
Approximate presence and absence of resource influx $I_\alpha$ and self-inhibition $1/K_\alpha$ is indicated by a tick or cross.
Refer to Fig.~\ref{fig:2} for details on how the diagrams was generated.
\textbf{(B)} Phase diagram for the hybrid model as a function of reciprocity and external resource influx $I_\alpha$, self-inhibition $1/K_\alpha=1$.
\textbf{(C)} 
(i) Example binned distribution of resource sensitivities to changes in their total depletion rate $C_\alpha$, sampled at $I_\alpha = 10^{-5}$, number of bins $=15$.
The green point is the sensitivity of near extinction resources $(C_\alpha \approx b_\alpha \approx 1)$  used in plot C ii.
(ii) Sensitivity of near-extinct resources to consumption (similar to (i)) as a function of influx rate along the stability-chaos boundary (Appendix D for details).
The detection threshold for peaks in sensitivity was set to 0.01 (dashed grey vertical line), so at $I_\alpha = 10^{-2}$ there was no detectable peak.
Points indicate average over realizations, error bars are 95\% confidence intervals for $n=80$ communities.
}}}
\label{fig:4}
\end{figure*}

\vskip 5pt
\label{results_analytics}\subsection*{Unified stability condition}To understand what drives the difference in dynamical regimes when changing resource supply mode, we used the cavity method to derive a unified stability condition for consumer--resource models with an arbitrary resource supply function $\mathcal{H}_\alpha(R_\alpha)$ (SI Appendices A--B). The cavity method involves deriving self-consistency equations for the steady-state abundance distributions of consumers and resources~\cite{bunin_ecological_2017,blumenthal_phase_2024,advani_statistical_2018,arnoulx_de_pirey_many-species_2024}. When the moments of the distributions corresponding to species and resource responses to perturbations diverge, communities become unstable. Applying this to the general class of consumer--resource models in Eqs.~\eqref{eq:res_dynamics}--\eqref{eq:sp_dynamics}, we obtain the following unified stability condition:

\begin{eqnarray}
    \rho^2 > \frac{S^*}{M} \frac{\langle \chi_{R}^2 \rangle}{\langle \chi_{R} \rangle^2},
    \label{eq:universal_stability_condition}
\end{eqnarray}
where  $S^*$ is the  number of surviving consumers, and $\chi_R = \{-\partial R_\alpha / \partial b_\alpha\}$ is the distribution of resource self-susceptibilities describing how a resource's abundance responds to perturbations in its growth or dilution rate.

To illustrate how this condition manifests for both models, we consider the simple limit of homogeneous resources ($\sigma_b=\sigma_K=\sigma_I=0$).
In this limit, the key difference between the two models is that self-renewing resources can go extinct ($M^*\leq M$ resources survive), while externally-supplied resources cannot go extinct because of the constant influx $I_\alpha\neq0$.
As a result, the ratio $\langle \chi_{R}^2 \rangle / \langle \chi_{R} \rangle^2$ simplifies to $M/M^*$ for self-renewing, and $1$ for externally-supplied resources (SI Appendix B), and the stability conditions reduce to:

\begin{eqnarray}
    \text{self-renewing:} \quad &\rho^2 > \frac{S^*}{M^*}, 
    \label{eq:stab_cond_homo_self} \\
    \text{externally-supplied:} \quad &\rho^2 > \frac{S^*}{M}.
    \label{eq:stab_cond_homo_ext}
\end{eqnarray}

Since ecosystems with externally-supplied resources are typically less niche-packed than self-renewing resources ($S^*/M\leq0.5$ for externally-supplied, $S^*/M^*\leq1$ for self-renewing~\cite{cui_effect_2020}), Eqs.~\eqref{eq:stab_cond_homo_self}--\eqref{eq:stab_cond_homo_ext} help explain that externally-supplied resources should typically support a larger stable region.
Note that our unified condition recovers the known stability condition for self-renewing resources, Eq.~\eqref{eq:stab_cond_homo_self}~\cite{blumenthal_phase_2024,liu_complex_2025}.

When resources are heterogeneous, Eq.~\eqref{eq:stab_cond_homo_ext} no longer holds and we must instead obtain $\langle \chi_{R}^2 \rangle / \langle \chi_{R} \rangle^2$ using the self-consistency equations (SI Appendices A--B).
This results in a more general stability condition for externally-supplied resources, which is rather different from that for self-renewing resources (Eq.~\eqref{eq:stab_cond_homo_self}):

\begin{eqnarray}
    \rho^2 > 
    \frac{M}{4 S^*}
        \left \langle 
        \left[
        1 - \frac{b_\alpha^{\rm eff}}{
            \sqrt{(b_\alpha^{\rm eff})^2 + \frac{4 S^* I_\alpha}{M \langle \chi_R \rangle}}}
        \right]^2
        \right \rangle,
        \label{eq:externally_supplied_condition}
\end{eqnarray}
where $b_\alpha^{\rm eff} = b_\alpha + \sum_i^{S^*} c_{i \alpha} N_i$ is the effective depletion rate of resource $\alpha$ due to outflux and consumption. Eq.~\eqref{eq:externally_supplied_condition} agrees remarkably well with our numerical phase diagrams (SI Appendix E Fig. S1 A--B).
In sum, our unified condition shows that stability depends not just on the interplay between reciprocity and niche packing~\cite{blumenthal_phase_2024,liu_complex_2025}, but also importantly on resource susceptibilities, specifically whether resources can go extinct. This difference in resource susceptibilities is key to why changing resource supply mode can drastically change the stable region.

\vskip 5pt
\label{results_influx}\subsection*{External resource influx suppresses chaos}Once unstable,  the externally-supplied resource model does not exhibit chaos. To understand what drives this loss of chaos, we investigated a hybrid model combining the two resource supply functions in Eqs.~\eqref{eq:self-renewing_supply}--\eqref{eq:externally-supplied_supply}:

\begin{equation}
    \mathcal{H}_\alpha (R_\alpha) = I_\alpha + R_\alpha \left[ b_\alpha - \frac{R_\alpha}{K_\alpha} \right],
\end{equation}
where $I_\alpha$ is the externally-supplied influx rate, $b_\alpha$ can be the intrinsic resource growth rate or dilution rate (depending on its sign), and $K_\alpha^{-1}$ is the strength of resource self-inhibition.
The hybrid model could describe, e.g., the resource-consumer dynamics of bacteria-phages in sewers, with bacteria supplied by both self-renewal and external inflow.
External influx  ($I_\alpha$) is unique to the externally-supplied model, while self-inhibition ($K_\alpha^{-1}$) is unique to the self-renewing model. 
Thus, tuning these terms enables the hybrid model to interpolate between the two supply functions.

We simulated phase diagrams for this model spanning four cases: in the presence or absence of influx and self-inhibition (Fig.~\ref{fig:4} A; SI Appendix D for details).
Without resource influx, the hybrid model had a small or no stable phase and a large chaotic phase, irrespective of self-limitation (Fig. \ref{fig:4} A, left two panels). 
On the other hand, with resource influx, the stable phase expanded and chaos was suppressed (Fig. \ref{fig:4} A, right two panels).
To quantitatively observe the loss of chaos, we performed additional simulations in which we systematically increased the influx rate.
Doing so systematically broadened the stable phase and suppressed both chaotic and infeasible phases, but the chaotic phase was suppressed faster, disappearing entirely at  $I_\alpha \geq 10^{-2}$ (Fig. \ref{fig:4} B). 
Analytic cavity calculations of the stability condition for the hybrid model were consistent with our unified stability condition Eq.~\eqref{eq:universal_stability_condition} (SI Appendix C). 
When evaluated explicitly, this condition resembled that of the externally-supplied model (SI Appendix C). This shows that external resource influx is a stronger driver of phase transitions and chaos than resource self-inhibition.

We now explain how increasing influx leads to loss of chaos. 
In consumer-resource models, a resource is driven towards its baseline abundance when it is heavily consumed (with total consumption $ C_\alpha = \sum_i^S c_{i \alpha} N_i \approx 1$), with the baseline set by the resource influx rate.
Thus, when there is no influx the baseline is zero, so resources can go extinct.
In the chaotic phase, near-baseline resources are highly sensitive to changes in their total depletion, which is typically quantified by the second moment of the response ~\cite{arnoulx_de_pirey_critical_2025}, and in this case given by $\langle(\mathrm{d} R_\alpha / \mathrm{d} C_\alpha )^2\rangle_{C_\alpha\approx b_a}$ (Fig. \ref{fig:4} C i).

One might na\"ively expect that raising influx will not affect chaos, since increasing the baseline resource abundance supports larger consumer populations and hence greater depletion.
However, as the rate of resource influx increases, resource concentrations become controlled more by influx than by consumption. 
Consequently, resources that are close to extinction may no longer be highly sensitive to perturbations such as consumer depletion. 
To test this, we varied resource influx in the hybrid model and measured the sensitivity of near-extinction resources $\langle(\mathrm{d} R_\alpha / \mathrm{d} C_\alpha )^2 \rangle_{C_\alpha \approx b_\alpha}$ near the stability boundary (see SI Appendix D for details).
We found that increasing resource influx decreased the sensitivity of near-extinct resources by several orders of magnitude. Sensitivity became practically non-detectable ($\ll 1$) at $I_\alpha \geq 10^{-2}$ (Fig. \ref{fig:4} C ii), exactly the influx rate at which we observed the loss of the chaotic phase (Fig. \ref{fig:4} B).
This confirmed that external resource influx suppresses chaos by reducing the sensitivity of resources to consumer depletion. 

\vskip 5pt
\label{discussion} 
\section*{Discussion}Our central finding is that the mode of resource supply can fundamentally alter the dynamical collective phases of an ecosystem.
While self-renewing resources yield stable, chaotic, and infeasible phases, externally-supplied resources yield only stable and infeasible phases, without chaos.
The unified stability condition shows that this difference arises because resources can go extinct under self-renewing but not external resource supply. 
External supply then suppresses chaos by desensitising near-extinction resources to perturbations. 

Our results contrast with previous work suggesting that resource supply does not alter ecosystem phase transitions~\cite{liu_complex_2025, liu_ecosystem_2024}.
Those studies prescribed fixed points with a chosen set of surviving species and resources, rather than letting them emerge from ecological dynamics.
As a result, they likely also implicitly set the resource susceptibilities to be the same for both forms of resource supply. 
Thus it might appear that resource supply has no effect on ecosystem stability.
Further work is needed to establish the source of the observed discrepancy. 


Recent experiments show microbial communities exhibiting persistent fluctuations under external resource supply \cite{hu_emergent_2022, gore_transition_2025, fernandez-gonzalez_microbial_2016}, yet our theory predicts suppression of chaos.
This suggests that the fluctuations observed in these experiments might arise from mechanisms other than resource-mediated interactions. Examples include pH modification~\cite{ratzke_strength_2020}, direct interactions~\cite{hu_emergent_2022, arnoulx_de_pirey_many-species_2024, arnoulx_de_pirey_critical_2025, mallmin_chaotic_2024,odwyer_whence_2018,mustri_accuracy_2025}, and periodic resource supply (serial dilution) in experiments versus constant resource supply (chemostats).
Identifying which of these mechanisms underlie chaos is key to controlling dynamical phases in natural and laboratory communities.

Finally, our results highlight that resource supply mode can fundamentally alter feasible phases of ecosystem dynamics. 
Our unified stability condition could help classify other modes of resource supply, e.g., cross-feeding, by the dynamical phases they support (SI Appendix B).
Analysing the chaotic phase using dynamical mean-field theory~\cite{arnoulx_de_pirey_critical_2025, arnoulx_de_pirey_many-species_2024, roy_numerical_2019} would also shed further light on the quantitative patterns of dynamical phases under different resource supply modes.
These directions would lead to a better understanding of different dynamical behaviours observed in natural ecosystems.


\vskip 5pt
\section*{Acknowledgements}
\label{acknowledgements} 
We thank G. Bunin, R. Mallikarjun, S. Pollak, P. Mehta,  and E. Blumenthal for valuable discussions. This research was supported in part by grant NSF PHY-2309135 to the Kavli Institute for Theoretical Physics (KITP) and the Gordon and Betty Moore Foundation Grant No. 2019.02. A.G. acknowledges support from the DAE under project no. RTI4001, an Ashok and Gita Vaish Junior Researcher Award, as well as the Centre for Artificial Learning and Intelligence for Biological Research and Education, ICTS-TIFR. 
W.S. acknowledges support from the Royal Society Wolfson Fellowship, as well as an Academy of Medical Sciences Professorship.
\pagebreak

\setcounter{figure}{0}
\renewcommand{\thefigure}{S\arabic{figure}}
\part*{Supplementary Information}

\appendix

\tableofcontents

\newpage

\section{Stability condition for externally-supplied resources using the cavity method}

\subsection{Steady-state community before introducing the cavity consumer and resource}

We begin with a steady-state community containing $S$ consumers and $M$ resources before the cavity consumer and resource $0$ are introduced.
The steady-state abundances of resident consumers and resources are described by the following equations: 

\begin{align}
\begin{split}
    \frac{\mathrm{d}N_{i\setminus 0}}{\mathrm{d}t} & = 0 
    = N_{i\setminus 0} \Biggl[ \sum_{\alpha = 1}^M g_{i \alpha} R_{\alpha\setminus 0} - d_i \Biggl].
\end{split}
\label{eq:pre-perturb consumer}
\end{align}

\begin{align}
\begin{split}
     \frac{\mathrm{d}R_{\alpha \setminus 0}}{\mathrm{d}t} &= 0
    = I_\alpha - b_\alpha R_{\alpha \setminus 0} - R_{\alpha \setminus 0} \sum_{i=1}^S c_{i \alpha} N_{i\setminus 0}.
\end{split}
\label{eq:pre-perturb resources}
\end{align}

Next, we substitute the growth and consumption coefficients with their parameter sampling rules (Main text, Parametrisation).
Starting with the resident consumer, we obtain

\begin{align}
\begin{split}
    0 & = N_{i\setminus 0} \Biggl[ \sum_{\alpha = 1}^M 
    \Big[ \frac{\mu_g}{M} + \frac{\sigma_g}{\sqrt{M}} z_{g, i \alpha} \Big] R_{\alpha\setminus 0} -
    d_i \Biggl]
    = N_{i\setminus 0} \Biggl[ 
    \mu_g \langle R \rangle
    + \frac{\sigma_g}{\sqrt{M}} \sum_{\alpha = 1}^M z_{g, i \alpha} R_{\alpha\setminus 0} - d_i \Biggl].
\end{split}
\label{eq:pre-perturb consumer mean fluct}
\end{align}

Note that $\mu_g \langle R \rangle = [\mu_g / M] \sum_{\alpha = 1}^M R_{\alpha\setminus 0}$, where $\langle R \rangle$ is the average abundance of resident resources, is an expression we define self-consistently later.

Repeating the same process with resident resources gives us

\begin{align}
\begin{split}
    0 &=
     I_\alpha - b_\alpha R_{\alpha \setminus 0}
     - R_{\alpha \setminus 0} \left[
      \mu_c \gamma^{-1} \langle N \rangle
      + \frac{\sigma_c}{\sqrt{M}} \sum_{i=1}^S \Big[ \rho z_{g, i \alpha} + \sqrt{1 - \rho^2} z_{c, i \alpha }\Big] N_{i\setminus 0}.
      \right].
\end{split}
\label{eq:pre-perturb resources mean fluct}
\end{align}

Note that $\mu_c \gamma^{-1} \langle N \rangle = [\mu_c / M] \sum_{i = 1}^S N_{i\setminus 0}$, where $\langle N \rangle$ is the average abundance of resident consumers, is an expression we define self-consistently later.

\subsection{Introducing the cavity consumer and resource}

Next, we invade the existing community with the cavity consumer and resource, which are statistically identical to the resident consumers and resources. 
The steady-state abundances for the resident consumers after invasion $N_i$ and resources $R_\a$ thus obey the following equations:

\begin{align}
\begin{split}
    \frac{\mathrm{d}N_i}{\mathrm{d}t} = 0
    &= N_i
    \Big[ 
    \sum_{\alpha = 1}^M g_{i \alpha} R_\alpha
    - d_i
    + g_{i 0} R_0
     \Biggl]
     \\
     & = N_i \Big[ 
    \mu_g \langle R \rangle + \frac{\sigma_g}{\sqrt{M}} \sum_{\alpha = 1}^M z_{g, i \alpha} R_\alpha
    - d_i
    + \frac{\sigma_g}{\sqrt{M}} z_{g, i 0} R_0
     \Biggl].
\end{split}
\label{eq: perturbed existing consumer rough mean fluct}
\end{align}

The effect of the cavity resource on the dynamics of consumer $i$ --- $\sigma_g / \sqrt{M} \sum_{\alpha = 1}^M z_{g, i 0} R_0$ --- can be interpreted as a small perturbation to the death rate of consumer $i$, $\Delta d_i$:

\begin{align}
    d_i \xrightarrow[]{\text{perturb}} d_i - \frac{\sigma_g}{\sqrt{M}} z_{g, i 0} R_0.
\label{eq: perturbed death rate}
\end{align}

Similarly, the new steady-state abundances of resident resources after invasion by the cavity variables obey the following equations:

\begin{align}
\begin{split}
    \frac{\mathrm{d}R_\alpha}{\mathrm{d}t} &= 0 = 
    I_\alpha - b_\alpha R_\alpha
    - R_\alpha \left[
    \mu_c \gamma^{-1} \langle N \rangle
    + \frac{\sigma_c}{\sqrt{M}} \sum_{i=1}^S \Big[ \rho z_{g, i \alpha} + \sqrt{1 - \rho^2} z_{c, i \alpha }\Big] N_i
    \right]
    - \frac{\sigma_c}{\sqrt{M}} R_\alpha \Big[ \rho z_{g, 0 \alpha} + \sqrt{1 - \rho^2} z_{c, 0 \alpha }\Big] N_0.
\end{split}
\label{eq: perturbed existing resources}
\end{align}

This effect of the cavity consumer on the dynamics of resource $\a$ can be interpreted as a small perturbation to the outflux rate of resource $\a$, $\Delta b_\a$:

\begin{align}
    b_\a \xrightarrow[]{\text{perturb}} b_\a + \frac{\sigma_c}{\sqrt{M}} \Big[ \rho z_{g, 0 \alpha} + \sqrt{1 - \rho^2} z_{c, 0 \alpha }\Big] N_0.
\label{eq: perturbed outflux rate}
\end{align}

We can relate the post-perturbation abundance of consumer $i$, $N_i$ with its pre-perturbation abundance $N_{i\wo0}$ using a linear response approximation, given by:

\begin{align}
\begin{split}
    N_i & \approx 
    N_{i\setminus 0} + \sum_{j=1}^S 
    \underbrace{\frac{\partial N_i}{\partial d_j}}_{v_{N, ij}} \Delta d_j
    + \sum_{\beta=1}^M 
    \underbrace{\frac{\partial N_i}{\partial b_\beta}}_{- \chi_{N,  i \beta}} \Delta b_\b.
    \\ &
    \approx N_{i\setminus 0} 
    - \sum_{j=1}^S v_{N, ij} \left[ \frac{\sigma_g}{\sqrt{M}} z_{g, j 0} R_0 \right]
    - \sum_{\beta=1}^M \chi_{N,  i \beta} \left[ \frac{\sigma_c}{\sqrt{M}} \Big[ \rho z_{g, 0 \beta} + \sqrt{1 - \rho^2} z_{c, 0 \beta} \Big] N_0 \right],
    \\ 
    & \text{where } v_{N, i j} = \frac{\partial N_i}{\partial d_j} \text{ and } \chi_{N, i \beta} = - \frac{\partial N_i}{\partial b_\beta},
\end{split}
\label{eq:taylor consumer}
\end{align}

where $v_{N, ij}$ and $\chi_{N,  i \beta}$ are susceptibility matrices characterising the response of the abundance of consumer $i$ to perturbations in the death rate of consumer $j$ and outflux rate of resource $\b$, respectively. 

Repeating the same process with resource $\alpha$:

\begin{align}
\begin{split}
    R_\alpha & \approx 
    R_{\alpha \setminus 0} + \sum_{j=1}^S 
    \underbrace{\frac{\partial R_\alpha}{\partial d_j}}_{v_{R,  \alpha j}} \Delta d_j
    + \sum_{\beta=1}^M 
    \underbrace{\frac{\partial R_\alpha}{\partial b_\beta}}_{-\chi_{R,  \alpha \beta}} \Delta b_\b
    \\ &
    \approx R_{\alpha \setminus 0} - \sum_{j = 1}^S v_{R,  \alpha j}
   \left[ \frac{\sigma_g}{\sqrt{M}} z_{g, j 0} R_0 \right]
    - \sum_{\beta=1}^M \chi_{R,  \alpha \beta} \left[ \frac{\sigma_c}{\sqrt{M}} \Big[ \rho z_{g, 0 \beta} + \sqrt{1 - \rho^2} z_{c, 0 \beta} \Big] N_0 \right],
    \\ 
    & \text{where } v_{R, \alpha j} = \frac{\partial R_\alpha}{\partial d_j} \text{ and } \chi_{R, \alpha \beta} = - \frac{\partial R_\alpha}{\partial b_\beta},
\end{split}
\label{eq:taylor resources}
\end{align}

where $v_{R,  \alpha j}$ and $\chi_{R,  \alpha \beta}$ are the susceptibility matrices characterising the response of the abundance of resource $\a$ to small perturbations in the death rate of consumer $j$ and supply rate of resource $\b$, respectively. 

Below, we collect all the four susceptibility matrices defined for the model for reference later, as:

\begin{align}
    v_{N, i j} = \frac{\partial N_i}{\partial d_j}, \quad
    v_{R, \alpha j} = \frac{\partial R_{\alpha}}{\partial d_j}, \quad
    \chi_{N, i \beta} = - \frac{\partial N_i}{\partial b_\beta}, \quad
    \chi_{R, \alpha \beta} = - \frac{\partial R_\alpha}{\partial b_\beta}.
\label{eq:susceptibilities}
\end{align}

\subsection{Self-consistent dynamics of the cavity consumer and resource}

The steady-state condition for the cavity consumer (with abundance $N_0$) and resource (with abundance $R_0$) are 

\begin{align}
\begin{split}
    \frac{\mathrm{d}N_0}{\mathrm{d}t} & = 0
    = N_0 \left[ 
    \mu_g \langle R \rangle 
    + \frac{\sigma_g}{\sqrt{M}} \sum_{\alpha=1}^M z_{g, 0 \alpha} R_\alpha
    + \frac{\sigma_g}{\sqrt{M}} z_{g, 0 0} R_0 
    - d_0 \right]
\end{split}
\label{eq:cavity consumer ode}
\end{align}

and

\begin{align}
\begin{split}
    \frac{\mathrm{d}R_0}{\mathrm{d}t} = 0 & = 
    I_0 - b_0 R_0
    - R_0 \left[
    \mu_c \gamma^{-1} \langle N \rangle
    + \frac{\sigma_c}{\sqrt{M}} \sum_{i=1}^S \left[ \rho z_{g, i 0} + \sqrt{1 - \rho^2} z_{c, i 0} \right] N_i
    +  \frac{\sigma_c}{\sqrt{M}} \left[ \rho z_{g, 0 0} + \sqrt{1 - \rho^2} z_{c, 0 0} \right] N_0
    \right].
\end{split}
\label{eq:cavity resource ode}
\end{align}

\subsubsection{Cavity consumer equation}

We first solve for the steady-state abundance of the cavity consumer $N_0$. 
Substituting expressions for $N_i$ and $R_\alpha$ from eq. \eqref{eq:taylor resources}, and plugging in the parameter sampling rules (Main text, Parametrisation) for $d_0$, we get:

\begin{align}
\begin{split}
    0 = N_0  \Biggl[ &
    \mu_g \langle R \rangle - \mu_d
    + \frac{\sigma_g}{\sqrt{M}} \sum_{\alpha=1}^M z_{g, 0 \alpha}  R_{\alpha \setminus 0}
    - \frac{\sigma_g^2}{M} R_0 \sum_{\alpha=1}^M \sum_{j = 1}^S v_{R,  \alpha j} z_{g, 0 \alpha} z_{g, j 0}
    \\ &
    - \frac{\sigma_g \sigma_c}{M} N_0 \sum_{\alpha=1}^M \sum_{\beta=1}^M \chi_{R,  \alpha \beta} 
        z_{g, 0 \alpha} \Big[ \rho z_{g, 0 \beta} + \sqrt{1 - \rho^2} z_{c, 0 \beta} \Big]
    + \frac{\sigma_g}{\sqrt{M}} z_{g, 0 0} R_0 
    - \sigma_d z_{d, 0}
    \Biggl].
\end{split}
\label{eq:cavity consumer taylor plugged in expanded}
\end{align}

To solve this equation and obtain the steady-state abundance $N_0$, we invoke the central limit theorem to express the many sums of weakly-correlated variables in eq. \eqref{eq:cavity consumer taylor plugged in expanded} as a single Gaussian random variable.
The Gaussianity of the sums will appear due to the central limit theorem irrespective of whether the underlying $c_{i\a}$ follow uniform, Bernoulli or Gaussian distributions. 
To express these sums as a single Gaussian variable, we need to compute its first and second moments (which will contribute to the mean and fluctuations) in terms of the non-zero $M$-leading-order moments of the sums in eq. \eqref{eq:cavity consumer taylor plugged in expanded}.

Most first moments in \eqref{eq:cavity consumer taylor plugged in expanded} are zero.
The only non-zero term that contributes to the first moment is

\begin{align}
\begin{split}
    & \left \langle
    - \frac{\sigma_g \sigma_c}{M} N_0 \sum_{\alpha=1}^M \sum_{\beta=1}^M \chi_{R,  \alpha \beta} 
        z_{g, 0 \alpha} \Big[ \rho z_{g, 0 \beta} + \sqrt{1 - \rho^2} z_{c, 0 \beta} \Big]
    \right \rangle    
    = - \sigma_g \sigma_c \rho \langle \chi_{R} \rangle N_0, 
\end{split}
\label{eq:cavity consumer mean term 4 part 1}
\end{align}

where $\langle \chi_{R} \rangle = 1/M\sum_{\alpha, \beta}^M \chi_{R, \alpha \beta} = 1/M \ \text{Trace}(\chi_{R, \alpha \beta})$ in the large $M$ limit is the average diagonal entry of the susceptibility matrix $\chi_{R, \a\b}$. 

The non-zero, non-vanishing second moments are 

\begin{align}
    \left \langle [-\sigma_d z_{0, d}]^2 \right \rangle = \sigma_d^2.
\label{eq:cavity consumer fluctuations term 4}
\end{align}

and

\begin{align}
\begin{split}
    \left \langle \left[
    \frac{\sigma_g}{\sqrt{M}} \sum_{\alpha = 1}^M z_{g, 0 \alpha} R_{\alpha \setminus 0}
    \right]^2 \right \rangle
    = \sigma_g^2 \langle R^2 \rangle.    
\end{split}
\label{eq:cavity consumer fluctuations term 2 part 1}
\end{align}

Note that here we have introduced the mean of the square of the steady-state resource abundance $\langle R^2 \rangle = \frac{1}{M} \sum_{\a=1}^M R_\a^2$, another unknown which we must later determine self-consistently.

One can show that the second moments of the other terms in \eqref{eq:cavity consumer taylor plugged in expanded}, besides the final term, are sub-leading in $M$.
(For a detailed discussion see \cite{bunin_ecological_2017} for GLV models and \cite{blumenthal_phase_2024} for CRMs.

Combining the moments computed above,  we can write a simplified version of Eq. \eqref{eq:cavity consumer taylor plugged in expanded}:

\begin{align}
\begin{split}
    0 =& N_0 \Biggl[
    \mu_g \langle R \rangle - \mu_d - \sigma_g \sigma_c \rho \langle \chi_{R} \rangle N_0
    + \sqrt{\sigma_g^2 \langle R^2 \rangle + \sigma_d^2} Z_N
    \Biggl],
\end{split}
\label{eq:cavity consumer final ode}
\end{align}

where $Z_N$ is a standard normal Gaussian variable. 
We can solve this equation to compute the cavity consumer's abundance $(N_0)$ at steady state.
There are two solutions that satisfy eq. \eqref{eq:cavity consumer final ode}):
When $N_0 = 0$ (extinction), or $N_0 = [\mu_g \langle R \rangle - \mu_d + \sqrt{\sigma_g^2 \langle R^2 \rangle + \sigma_d^2} Z_N] / \sigma_g \sigma_c \rho \langle \chi_{R} \rangle$ (survival). 
This means the cavity consumer abundance follows a normal distribution truncated at 0:

\begin{align}
\begin{split}
    N_0 = & \text{max} \left \{ 0, \frac{\mu_g \langle R \rangle - \mu_d + \sqrt{\sigma_g^2 \langle R^2 \rangle + \sigma_d^2} Z_N}{\sigma_g \sigma_c \rho \langle \chi_{R} \rangle} \right \}.
\end{split}
\label{eq:cavity consumer distribution}
\end{align}

We assign the probability a cavity consumer will survive ($N_0 > 0$) as $S^*/S = \phi_N$, which we will later determine self-consistently.

\subsubsection{Cavity resource equation}

We can now repeat similar steps to solve Eq. \eqref{eq:cavity resource ode} to compute the steady-state abundance of resource 0.
Substituting our linear response approximation for $N_i$ from Eq. \eqref{eq:taylor consumer} in eq. \eqref{eq:cavity resource ode}, we get:

\begin{align}
\begin{split}
    0 = 
    \underbrace{\mu_I + \sigma_I z_{I, 0}}_{I_0} 
    - (\underbrace{\mu_b + \sigma_b z_{b, 0}}_{b_0}) R_0
    - R_0 \Bigg[ &
        \mu_c \gamma^{-1} \langle N \rangle
        \\ &
        + \frac{\sigma_c}{\sqrt{M}} \sum_{i=1}^S \left[ \rho z_{g, i 0} + \sqrt{1 - \rho^2} z_{c, i 0} \right]
        \Biggl[
        N_{i\setminus 0} 
        - \sum_{j=1}^S v_{N, ij} \left[ \frac{\sigma_g}{\sqrt{M}} z_{g, j 0} R_0 \right]
        \\ & \hphantom{ + \frac{\sigma_c}{\sqrt{M}} \sum_{i=1}^S \left[ \rho z_{g, i 0} + \sqrt{1 - \rho^2} z_{c, i 0} \right]
        \Big[}
        - \sum_{\beta=1}^M \chi_{N,  i \beta} \left[ \frac{\sigma_c}{\sqrt{M}} \Big[ \rho z_{g, 0 \beta} + \sqrt{1 - \rho^2} z_{c, 0 \beta} \Big] N_0 \right]
        \Biggl]
        \\ &
        + \frac{\sigma_c}{\sqrt{M}} \left[ \rho z_{g, 0 0} + \sqrt{1 - \rho^2} z_{c, 0 0} \right] N_0
    \Big].
\end{split}
\label{eq:cavity resource taylor added}
\end{align}

Similar to the cavity consumer, we need to compute the first and second non-vanishing moments of the terms in eq. \eqref{eq:cavity resource taylor added} to express these sums as simplified random variables.
Note that we cannot use the central limit theorem to combine the resource influx rate $I_0$ with consumption and dilution rates  because it is mulitplied by $R_0$, so it is not considered in the following calculations.

The only term with a non-zero, leading order first moment is

\begin{align}
\begin{split}
    & \left \langle 
    - R_0 \frac{\sigma_c}{\sqrt{M}} \sum_{i=1}^S \left[ \rho z_{g, i 0} + \sqrt{1 - \rho^2} z_{c, i 0} \right]
    \Biggl[ - \sum_{j=1}^S v_{N, ij} \left[ \frac{\sigma_g}{\sqrt{M}} z_{g, j 0} R_0 \right] \Biggl]
    \right \rangle 
    = \sigma_g \sigma_c \rho \gamma^{-1} \langle v_{N} \rangle R_0^2.
\end{split}
\label{eq:cavity resource mean incomplete}
\end{align}

where $\langle v_{N} \rangle = 1/S \sum_{i,j}^S v_{N, i j} = 1/S \ \text{Trace}(v_{N, i j})$ in the large $S$ limit is the average diagonal entry of the susceptibility matrix $v_{N, ij}$, and quantifies the average change in the abundance of consumer $i$ upon a small perturbation to its death rate.

The terms with non-zero, leading-order second moments are

\begin{align}
    \left \langle [- \sigma_b z_{b, 0} R_0]^2 \right \rangle = \sigma_b^2 R_0^2
\label{eq:cavity resource fluctuations influx}
\end{align}

and

\begin{align}
& \left \langle 
\left[ - R_0\frac{\sigma_c}{\sqrt{M}} \sum_{i=1}^S \left[ \rho z_{g, i 0} + \sqrt{1 - \rho^2} z_{c, i 0} \right]
        N_{i \setminus 0} \right]^2
    \right \rangle
    = 
    \sigma_c^2 \gamma^{-1} \langle N^2 \rangle R_0^2.
\label{eq:cavity resource fluctuations term 1 part 1}
\end{align}

Note that here we have introduced the mean of the square of the steady-state consumer abundance $\langle N^2\rangle = \frac{1}{S} \sum_{i=1}^S N_i^2$, another unknown which we must later determine self-consistently.

Similar to the case of the cavity consumer, one can show that the second moments of the other sums and the cavity consumer fluctuation are sub-leading compared to the above terms.

We can now write the dynamics of the cavity resource in terms of the non-vanishing moments.

\begin{align}
\begin{split}
    0 =&
    [\mu_I + \sigma_I z_{I, 0}]
    - \left[ \mu_b + \mu_c \gamma^{-1} \langle N \rangle + \sqrt{\sigma_c^2 \gamma^{-1} \langle N^2 \rangle + \sigma_b^2} Z_R \right] R_0 
    + \sigma_g \sigma_c \rho \gamma^{-1} \langle v_{N} \rangle R_0^2,
\end{split}
\label{eq:cavity resource final ode}
\end{align}

where the $-$ that would persist in $\sqrt{\sigma_c^2 \gamma^{-1} \langle N^2 \rangle + \sigma_b^2} Z_R$ has been absorbed into $Z_R$ for convenience.

Eq. \eqref{eq:cavity resource final ode} is a quadratic, meaning the solution for the cavity resource abundance $R_0$ is

\begin{align}
\begin{split}
    R_0 = & 
    \frac{
        \mu_b + \mu_c \gamma^{-1} \langle N \rangle 
        + \sqrt{\sigma_c^2 \gamma^{-1} \langle N^2 \rangle + \sigma_b^2} Z_R}{
        2 \sigma_c \sigma_g \rho \gamma^{-1} \langle v_{N} \rangle}
    \\ & \pm 
    \frac{\sqrt{
    [ \mu_b + \mu_c \gamma^{-1} \langle N \rangle + \sqrt{\sigma_c^2 \gamma^{-1} \langle N^2 \rangle + \sigma_b^2} Z_R ]^2 
    - 4[\sigma_c \sigma_g \rho \gamma^{-1} \langle v_{N} \rangle] [\mu_I + \sigma_I z_{I, 0}]
    }}{2 \sigma_c \sigma_g \rho \gamma^{-1}\langle v_{N} \rangle}.
\end{split}
\label{eq:cavity resource quadratic formula}
\end{align}

As we will show later on, $\langle v_{N} \rangle \leq 0$, so the only biologically relevant solution for $R_0$ is the negative solution.

Assuming that the means in the numerator of the second term in \eqref{eq:cavity resource quadratic formula} are $\gg$ than their fluctuations, we can approximate the second term by taking its linear binomial expansion about its mean:
This means $R_0$ can be approximated as

\begin{align}
\begin{split}
    R_0 \approx
    \frac{1}{2 \sigma_c \sigma_g \rho \gamma^{-1} \langle v_{N} \rangle}
    \Biggl[ &
    \mu_b + \mu_c \gamma^{-1} \langle N \rangle 
    + \sqrt{\sigma_c^2 \gamma^{-1} \langle N^2 \rangle + \sigma_b^2} Z_R
    - \sqrt{[\mu_b + \mu_c \gamma^{-1} \langle N \rangle]^2 
            - 4 \sigma_c \sigma_g \rho \gamma^{-1} \langle v_{N} \rangle \mu_I}
    \\ &
    - \frac{
        2 [\mu_b + \mu_c \gamma^{-1} \langle N \rangle] \sqrt{\sigma_c^2 \gamma^{-1} \langle N^2 \rangle + \sigma_b^2} Z_R
        + [\sigma_c^2 \gamma^{-1} \langle N^2 \rangle + \sigma_b^2] Z_R^2
        - 4 \sigma_c \sigma_g \rho \gamma^{-1} \langle v_{N} \rangle \sigma_I z_{I, 0}}{
        2 \sqrt{[\mu_b + \mu_c \gamma^{-1} \langle N \rangle]^2 
            - 4 \sigma_c \sigma_g \rho \gamma^{-1} \langle v_{N} \rangle \mu_I}}
    \Biggl].
\end{split}
\label{eq:cavity resource distribution}
\end{align}

\subsection{Self-consistency equations}
\label{box:sces}

Since the cavity species and resource are statistically identical to the resident species and resources respectively, we invoke self-averaging to claim that the distribution of all consumer and resources in the steady-state community follow the same distributions as the cavity consumer and resource.
Now, we derive the self-consistency equations describing the statistical moments of these abundance distributions: $\phi_N$, $\langle N \rangle$ and $\langle R \rangle$, $\langle N^2 \rangle$ and $\langle R^2 \rangle$, and $\langle v_{N} \rangle$ and $\langle \chi_{R} \rangle$.

\subsubsection{The statistical moments of the consumer abundance distribution}

Several of the unknowns are the moments of the consumer abundance distribution: the zeroth moment represents the consumer survival probability of consumer $(\phi_N)$, the first moments represent the average consumer abundance $\langle N \rangle$, and similarly for the second moment $\langle N^2 \rangle$ and $\langle R^2 \rangle$. 
The $j^{\text{th}}$ moment of the zero-truncated Gaussian describing the consumer abundance distribution is 

\begin{align}
\begin{split}
    & w_j \left( \mu_{g} \langle R \rangle - \mu_d, \sqrt{\sigma_g^2 \langle R^2 \rangle + \sigma_d^2} \right) = 
    \\ & \quad
    \left[ \frac{\sqrt{\sigma_g^2 \langle R^2 \rangle + \sigma_d^2}}{\sigma_g \sigma_c \rho \langle \chi_{R} \rangle} \right]^j
    \int_{- \frac{\mu_{g} \langle R \rangle - \mu_d}{\sqrt{\sigma_g^2 \langle R^2 \rangle + \sigma_d^2}}}^\infty 
    \frac{1}{\sqrt{ 2 \pi}}
    \exp \left( - \frac{z^2}{2} \right) 
    \left[ z + \frac{\mu_{g} \langle R \rangle - \mu_d}{\sqrt{\sigma_g^2 \langle R^2 \rangle + \sigma_d^2}} \right]^j
    dz.
\end{split}
\label{eq:consumer truncated gaussian}
\end{align}

The 0th moment $\phi_N$ is

\begin{align}
    \phi_N = 
    \int_{- \frac{\mu_{g} \langle R \rangle - \mu_d}{\sqrt{\sigma_g^2 \langle R^2 \rangle + \sigma_d^2}}}^\infty 
    \frac{1}{\sqrt{ 2 \pi}}
    \exp \left( - \frac{z^2}{2} \right) dz
    = \Phi \left( \frac{\mu_{g} \langle R \rangle - \mu_d}{\sqrt{\sigma_g^2 \langle R^2 \rangle + \sigma_d^2}} \right),
\label{eq:consumer 0th moment}
\end{align}
where $\Phi(...)$ is $1/2 \times$the complementary error function. 

The 1st moment $\langle N \rangle$ is

\begin{align}
\begin{split}
    \langle N \rangle  = 
    \frac{\sqrt{\sigma_g^2 \langle R^2 \rangle + \sigma_d^2}}{\sigma_g \sigma_c \rho \langle \chi_{R} \rangle}
    \left[\frac{1}{\sqrt{2 \pi}} \exp \left( -\frac{[\mu_g \langle R \rangle - \mu_d]^2}{2[\sigma_g^2 \langle R^2 \rangle + \sigma_d^2]} \right) + \frac{\mu_{g} \langle R \rangle - \mu_d}{\sqrt{\sigma_g^2 \langle R^2 \rangle + \sigma_d^2}} \ \Phi \left( \frac{\mu_{g} \langle R \rangle - \mu_d}{\sqrt{\sigma_g^2 \langle R^2 \rangle + \sigma_d^2}} \right) \right].
\end{split}
\label{eq:consumer 1st moment}
\end{align}

The 2nd moment $\langle N^2 \rangle$ is

\begin{align}
\begin{split}
     \langle N^2 \rangle =& 
    \left[ \frac{\sqrt{ \sigma_g^2 \langle R^2 \rangle + \sigma_d^2}}{\sigma_g \sigma_c \rho \langle \chi_{R} \rangle} \right]^2
    \Biggl[
    \frac{\mu_{g} \langle R \rangle - \mu_d}{\sqrt{ 2 \pi [\sigma_g^2 \langle R^2 \rangle + \sigma_d^2]} }
    \exp \left( -\frac{[\mu_g \langle R \rangle - \mu_d]^2}{2[\sigma_g^2 \langle R^2 \rangle + \sigma_d^2]} \right)
    \\ & \hphantom{\left[ \frac{\sqrt{ \sigma_g^2 \langle R^2 \rangle + \sigma_d^2}}{\sigma_g \sigma_c \rho \langle \chi_{R} \rangle} \right]^2 \Biggl[}
    + \left[ 1 + \frac{[\mu_g \langle R \rangle - \mu_d]^2}{\sigma_g^2 \langle R^2 \rangle + \sigma_d^2} \right] 
    \Phi \Bigg( \frac{\mu_{g} \langle R \rangle - \mu_d}{\sqrt{\sigma_g^2 \langle R^2 \rangle + \sigma_d^2}}  \Bigg) 
     \Biggl].
\end{split}
\label{eq:consumer 2nd moment}
\end{align}

\subsubsection{The statistical moments of the resource abundance distribution}

When resources are externally-supplied, their abundances do not follow a truncated Gaussian.
To obtain the average resource abundance ($\langle R \rangle$) and its fluctuations ($\langle R^2 \rangle$), we instead calculate the expected value of $R_0$ and $R_0^2$ using the expression for $R_0$ in eq. \eqref{eq:cavity resource distribution}.

The 1st moment, or average resource abundance $\langle R \rangle$, is

\begin{align}
\begin{split} 
    \langle R \rangle  = & 
    \frac{1}{2 \sigma_c \sigma_g \rho \gamma^{-1} \langle v_{N} \rangle}
    \Biggl[
    \mu_b + \mu_c \gamma^{-1} \langle N \rangle 
    - \sqrt{[\mu_b + \mu_c \gamma^{-1} \langle N \rangle]^2 
            - 4 \sigma_c \sigma_g \rho \gamma^{-1} \langle v_{N} \rangle \mu_I}
    \\ & \hphantom{\frac{1}{2 \sigma_c \sigma_g \rho \gamma^{-1} \langle v_{N} \rangle} \Biggl[}
    - \frac{\sigma_c^2 \gamma^{-1} \langle N^2 \rangle + \sigma_b^2}{
        2 \sqrt{[\mu_b + \mu_c \gamma^{-1} \langle N \rangle]^2 - 4 \sigma_c \sigma_g \rho \gamma^{-1} \langle v_{N} \rangle \mu_I}}
    \Biggl].
\end{split}
\label{eq:resource 1st moment}
\end{align}

The 2nd moment is

\begin{align}
\begin{split}
    \langle R^2 \rangle =
     \frac{1}{4 [ \sigma_g \sigma_c \rho \gamma^{-1} \langle v_{N} \rangle]^2} \Biggl[ &
     2 [\mu_b + \mu_c \gamma^{-1} \langle N \rangle]^2 
     + 2[\sigma_c^2 \gamma^{-1} \langle N^2 \rangle + \sigma_b^2] 
     - 4 \sigma_g \sigma_c \rho \gamma^{-1} \langle v_{N} \rangle \mu_I 
     \\ &
     - 2[\mu_b + \mu_c \gamma^{-1} \langle N \rangle] \sqrt{[\mu_b + \mu_c \gamma^{-1} \langle N \rangle]^2 - 4 \sigma_g \sigma_c \rho \gamma^{-1} \langle v_{N} \rangle \mu_I}
     \\ &
     - \frac{
        3[\mu_b + \mu_c \gamma^{-1} \langle N \rangle][\sigma_c^2 \gamma^{-1} \langle N^2 \rangle + \sigma_b^2]}{\sqrt{[\mu_b + \mu_c \gamma^{-1} \langle N \rangle]^2 
        - 4  \sigma_g \sigma_c \rho \gamma^{-1} \langle v_{N} \rangle \mu_I}} 
    \\ &
    + \frac{4 [\mu_b + \mu_c \gamma^{-1} \langle N \rangle]^2 [\sigma_c^2 \gamma^{-1} \langle N^2 \rangle + \sigma_b^2]
        + 3[\sigma_c^2 \gamma^{-1} \langle N^2 \rangle + \sigma_b^2]^2 
        + 16 [\sigma_g \sigma_c \rho \gamma^{-1} \langle v_{N} \rangle]^2 \sigma_I^2}{4 [[\mu_b + \mu_c \gamma^{-1} \langle N \rangle]^2 - 4  \sigma_g \sigma_c \rho \gamma^{-1} \langle v_{N} \rangle \mu_I]}
      \Biggl].
\end{split}
\label{eq:resource 2nd moment final}
\end{align}

\subsubsection{The susceptibilities}

The susceptibility $\langle v_{N} \rangle$, which we remind ourselves equals $\langle \partial N_i / d_i \rangle$ or equivalently $\langle \partial N_0 / \partial d_0 \rangle$, is given by

\begin{align}
    \langle v_{N} \rangle 
    & = 
    \left \langle 
    \frac{\partial}{\partial d_0} \left(
    \text{max} \left\{0, 
    \frac{\mu_g \langle R \rangle - \mu_d + \sqrt{\sigma_g^2 \langle R^2 \rangle + \sigma_d^2} Z_N}{\sigma_g \sigma_c \rho \langle \chi_{R} \rangle}
    \right\} \right) \right \rangle
    \\ & = 
    - \frac{\phi_N}{\sigma_g \sigma_c \rho \langle \chi_{R} \rangle}.
\label{eq:v_N}
\end{align}

The susceptibility $\langle \chi_{R} \rangle = - \langle \partial R_\a / \partial b_\a \rangle \equiv - \langle \partial R_0 / \partial b_0 \rangle $ is

\begin{align}
\begin{split}
    \langle \chi_{R} \rangle =& 
    - \Biggl \langle 
    \frac{\partial}{\partial b_0} \Biggl(
     \frac{
        \mu_b + \mu_c \gamma^{-1} \langle N \rangle 
        + \sqrt{\sigma_c^2 \gamma^{-1} \langle N^2 \rangle + \sigma_b^2} Z_R}{
        2 \sigma_c \sigma_g \rho \gamma^{-1} \langle v_{N} \rangle}
    \\ & \hphantom{Biggl \langle \frac{\partial}{\partial b_0} \Biggl(} - 
    \frac{\sqrt{
    [ \mu_b + \mu_c \gamma^{-1} \langle N \rangle + \sqrt{\sigma_c^2 \gamma^{-1} \langle N^2 \rangle + \sigma_b^2} Z_R ]^2 
    - 4[\sigma_c \sigma_g \rho \gamma^{-1} \langle v_{N} \rangle] [\mu_I + \sigma_I z_{I, 0}]
    }}{2 \sigma_c \sigma_g \rho \gamma^{-1}\langle v_{N} \rangle}
    \Biggl) \Biggl \rangle. 
    \\
    \approx&
    - \frac{1}{2 \sigma_c \sigma_g \rho \gamma^{-1} \langle v_{N} \rangle}
    \Biggl[
    1 
    - \frac{\mu_b + \mu_c \gamma^{-1} \langle N \rangle}{\sqrt{[\mu_b + \mu_c \gamma^{-1} \langle N \rangle]^2 
            - 4 \sigma_c \sigma_g \rho \gamma^{-1} \langle v_{N} \rangle \mu_I}}
    \\ & \hphantom{- \frac{1}{2 \sigma_c \sigma_g \rho \gamma^{-1} \langle v_{N} \rangle} \Biggl[}
    + \frac{3 [\mu_b + \mu_c \gamma^{-1} \langle N \rangle][\sigma_c^2 \gamma^{-1} \langle N^2 \rangle + \sigma_b^2]}{
    2 \left[[\mu_b + \mu_c \gamma^{-1} \langle N \rangle]^2 
            - 4 \sigma_c \sigma_g \rho \gamma^{-1} \langle v_{N} \rangle \mu_I\right]^{\frac{3}{2}}}
     \Biggl]       
\end{split}
\label{eq:chi_R}
\end{align}

\subsection{Stability condition}

Having derived properties of typical steady-state communities assembled from a pool of $S$ consumers and $M$ resources, we now proceed to derive when we expect these steady-states to be dynamically stable. 
For this, we will assess the sensitivity of the community steady-state to small perturbations in the abundances of all surviving consumers and resources at steady-state.
(By perturbing only surviving consumers and all resources, we assume that our steady state is uninvadable.)
When the distributions describing the sensitivities of consumers and resources to perturbations becomes ill-defined, the community becomes unstable.
We will find that it is the second moment that becomes ill-defined, and the stability boundary thus corresponds to the condition for this second moment to diverge.

\subsubsection{Sensitivity of consumer abundances to perturbations}

From eq. (\ref{eq:cavity consumer final ode}), the steady-state abundance of the surviving cavity species $N_0^+$ can be written as

\begin{align}
    N_0^+ & = \frac{1}{ \sigma_g \sigma_c \rho \langle \chi_{R} \rangle} \Biggl[
    \mu_g \langle R \rangle - \mu_d
    + \frac{\sigma_g}{\sqrt{M}} \sum_{\alpha = 1}^M z_{g, 0 \alpha} R_{\alpha \wo 0}
    - \sigma_d z_{d, 0}
    \Biggl],
\label{eq:surviving_consumer}
\end{align}

We now apply a random perturbation $\varepsilon \eta_\alpha$ to all resources $R_{\a \wo 0}$: $R_{\a\wo 0} \rightarrow R_{\a\wo 0} + \varepsilon \eta_\alpha$ --- where $\varepsilon\ll1$ and $\eta_\alpha$ is a standard normal variable.
After the perturbation, the steady-state abundance of the surviving cavity consumer is

\begin{align}
    N_0^+ & = \frac{1}{ \sigma_g \sigma_c \rho \langle \chi_{R} \rangle} \Biggl[
    \mu_g \langle R \rangle - \mu_d
    + \frac{\sigma_g}{\sqrt{M}} \sum_{\alpha = 1}^M z_{g, 0 \alpha} \left[ R_{\a\wo 0} + \varepsilon \eta_\alpha \right]
    - \sigma_d z_{d, 0}
    \Biggl].
\label{eq:surviving_consumer_perturbed}
\end{align}

Next, we compute the sensitivity of this consumer $d N_0^+ / d \varepsilon$ by taking its derivative with respect to the perturbation $\varepsilon$, given by:

\begin{align}
    \frac{\mathrm{d} N_0^+}{\mathrm{d} \varepsilon} & = 
    \frac{1}{ \sigma_g \sigma_c \rho \langle \chi_{R} \rangle} \Biggl[
    \frac{\sigma_g}{\sqrt{M}} \sum_{\alpha = 1}^M z_{g, 0 \alpha} \left[ \frac{\mathrm{d} R_{\a\wo 0}}{\mathrm{d} \varepsilon}  + \eta_\alpha \right]
    \Biggl].
\label{eq:dNde}
\end{align}

The first moment $\langle d N_0^+ / d \varepsilon \rangle = 0$.
However, this is not the case for the second moment $\langle [d N_0^+ / d \varepsilon]^2 \rangle$:

\begin{align}
    \left \langle \left[ \frac{\mathrm{d} N_0^+}{\mathrm{d} \varepsilon} \right]^2 \right \rangle 
    & = 
    \frac{1}{ \left[ \sigma_g \sigma_c \rho \langle \chi_{R} \rangle \right]^2} 
    \left \langle \Biggl[
    \frac{\sigma_g}{\sqrt{M}} \sum_{\alpha = 1}^M z_{g, 0 \alpha} \left[ \frac{\mathrm{d} R_{\a\wo 0}}{\mathrm{d} \varepsilon}  + \eta_\alpha \right]
    \Biggl]^2 \right \rangle
    \\ &=
    \frac{1}{\left[ \sigma_c \rho \langle \chi_{R} \rangle \right]^2}
    \left[ \left \langle \left[\frac{\mathrm{d}R_{\a\wo 0}}{d \varepsilon} \right]^2 \right \rangle + 1 \right.
\label{eq:dNde2}
\end{align}

Because the cavity resource is statistically identical to all community members, we can substitute $ \langle \left[ R_{\a\wo 0} / d \varepsilon \right]^2 \rangle$ with $ \langle \left[ R_0 / d \varepsilon \right]^2 \rangle$:

\begin{align}
    \left \langle \left[ \frac{\mathrm{d}N_0^+}{d \varepsilon} \right]^2 \right \rangle
    = & 
    \frac{1}{\left[ \sigma_c \rho \langle \chi_{R} \rangle \right]^2}
    \left[ \left \langle \left[\frac{\mathrm{d}R_0}{d \varepsilon} \right]^2 \right \rangle + 1 \right].
\label{eq:dNde2_cavity}
\end{align}

\subsubsection{Sensitivity of resource abundances to perturbations}

Starting with eq. \eqref{eq:cavity resource quadratic formula}, the cavity resource abundance $R_0$ can be described by the following equation:

\begin{align}
\begin{split}
    R_0 =
    \frac{1}{2 \sigma_c \sigma_g \rho \gamma^{-1} \langle v_{N} \rangle} \Biggl[ &
        \mu_b + \mu_c \gamma^{-1} \langle N \rangle 
         + \frac{\sigma_c}{\sqrt{M}} \sum_{i=1}^{S^*} \left[ \rho z_{g, i 0} + \sqrt{1 - \rho^2} z_{c, i 0} \right] {N_{i/0}^+} 
    + \sigma_b z_{b, 0}
    \\ & - 
    \Biggl[
    \Biggl[ \mu_b + \mu_c \gamma^{-1} \langle N \rangle
         + \frac{\sigma_c}{\sqrt{M}} \sum_{i=1}^{S^*} \left[ \rho z_{g, i 0} + \sqrt{1 - \rho^2} z_{c, i 0} \right] {N_{i/0}^+} 
        + \sigma_b z_{b, 0} \Biggl]^2 
    \\ & \hphantom{- \Biggl[ \Biggl[}
    - 4 \sigma_c \sigma_g \rho \gamma^{-1} \langle v_{N} \rangle [\mu_I + \sigma_I z_{I, 0}]
    \Biggl]^{\frac{1}{2}} \Biggl].
\end{split}
\label{eq:resource}
\end{align}

We now perturb surviving consumers abundances in a similar fashion to surviving resources in eq. \eqref{eq:surviving_consumer_perturbed} -- $N_{i\wo0}^+ \rightarrow N_{i\wo0}^+ + \varepsilon \eta_i$ --- giving us

\begin{align}
\begin{split}
    R_0 = 
    \frac{1}{2 \sigma_c \sigma_g \rho \gamma^{-1} \langle v_{N} \rangle} \Biggl[ &
        \mu_b + \mu_c \gamma^{-1} \langle N \rangle 
         + \frac{\sigma_c}{\sqrt{M}} \sum_{i=1}^{S^*} \left[ \rho z_{g, i 0} + \sqrt{1 - \rho^2} z_{c, i 0} \right] \left[ {N_{i/0}^+} +  \varepsilon \eta_i \right]
    + \sigma_b z_{b, 0}
    \\ & - 
    \Biggl[
    \Biggl[ \mu_b + \mu_c \gamma^{-1} \langle N \rangle
         + \frac{\sigma_c}{\sqrt{M}} \sum_{i=1}^{S^*} \left[ \rho z_{g, i 0} + \sqrt{1 - \rho^2} z_{c, i 0} \right] \left[ {N_{i/0}^+} +  \varepsilon \eta_i \right] 
        + \sigma_b z_{b, 0} \Biggl]^2 
    \\ & \hphantom{- \Biggl[ \Biggl[}
    - 4 \sigma_c \sigma_g \rho \gamma^{-1} \langle v_{N} \rangle [\mu_I + \sigma_I z_{I, 0}]
    \Biggl]^{\frac{1}{2}} \Biggl].
\end{split}
\label{eq:resource_perturbed}
\end{align}

The cavity resource's sensitivity to perturbations is

\begin{align}
\begin{split}
    \frac{ \mathrm{d} R_0}{\mathrm{d} \varepsilon} =
    \frac{1}{2 \sigma_c \sigma_g \rho \gamma^{-1} \langle v_{N} \rangle} \Biggl[ &
    \frac{\sigma_c}{\sqrt{M}} \sum_{i=1}^{S^*} \left[ \rho z_{g, i 0} + \sqrt{1 - \rho^2} z_{c, i 0} \right] \left[ \frac{ \mathrm{d} {N_{i/0}^+}}{\mathrm{d} \varepsilon} + \eta_i \right]
    \\ & 
    - \frac{ \mathrm{d}}{\mathrm{d} \varepsilon} \Biggl[
    \Biggl[ \mu_b + \mu_c \gamma^{-1} \langle N \rangle
         + \frac{\sigma_c}{\sqrt{M}} \sum_{i=1}^{S^*} \left[ \rho z_{g, i 0} + \sqrt{1 - \rho^2} z_{c, i 0} \right] \left[ {N_{i/0}^+} +  \varepsilon \eta_i \right] 
        + \sigma_b z_{b, 0} \Biggl]^2 
    \\ & \hphantom{- \frac{ \mathrm{d}}{\mathrm{d} \varepsilon} \Biggl[ \Biggl[}
    - 4 \sigma_c \sigma_g \rho \gamma^{-1} \langle v_{N} \rangle [\mu_I + \sigma_I z_{I, 0}]
    \Biggl]^{\frac{1}{2}} \Biggl].
\end{split}
\label{eq:R0 differentiate with respect to e}
\end{align}

To differentiate the last term of eq. \eqref{eq:R0 differentiate with respect to e}, we denote $f(\varepsilon) = \mu_b + \mu_c \gamma^{-1} \langle N \rangle + [\sigma_c / {\sqrt{M}}] \sum_{i=1}^{S^*} [ \rho z_{g, i 0} + \sqrt{1 - \rho^2} z_{c, i 0} ] [ {N_{i/0}^+} +  \varepsilon \eta_i ] + \sigma_b z_{b, 0}$.
Therefore, the final term can be reduced to 

\begin{align}
\begin{split}
    - \frac{\mathrm{d}}{\mathrm{d} \varepsilon} [f(\varepsilon)^2 - 4 \sigma_c \sigma_g \rho \gamma^{-1} \langle v_{N} \rangle [\mu_I + \sigma_I z_{I, 0}]]^{\frac{1}{2}}
=& 
- \frac{1}{2 [f(\varepsilon)^2 - 4 \sigma_c \sigma_g \rho \gamma^{-1} \langle v_{N} \rangle [\mu_I + \sigma_I z_{I, 0}]]^{\frac{1}{2}}}
        \times 2 f(\varepsilon) \frac{\mathrm{d} f(\varepsilon)}{\mathrm{d} \varepsilon}
\\ =&
- \frac{
    f(\varepsilon) \sigma_c \sum_{i=1}^{S^*} \left[ \rho z_{g, i 0} + \sqrt{1 - \rho^2} z_{c, i 0} \right] 
    \left[ \frac{ \mathrm{d} {N_{i/0}^+}}{\mathrm{d} \varepsilon} + \eta_i \right]
    }{\sqrt{M} [f(\varepsilon)^2 - 4 \sigma_c \sigma_g \rho \gamma^{-1} \langle v_{N} \rangle [\mu_I + \sigma_I z_{I, 0}]]^{\frac{1}{2}}}
\end{split}
\label{eq:dRde last term differentiate chain rule}
\end{align}

Therefore, $\mathrm{d} R_0 / \mathrm{d} \varepsilon$ is 

\begin{align}
\begin{split}
    \frac{ \mathrm{d} R_0}{\mathrm{d} \varepsilon} =
    \frac{1}{2 \sigma_c \sigma_g \rho \gamma^{-1} \langle v_{N} \rangle} & \Biggl[
    \frac{\sigma_c}{\sqrt{M}} \sum_{i=1}^{S^*} \left[ \rho z_{g, i 0} + \sqrt{1 - \rho^2} z_{c, i 0} \right] \left[ \frac{ \mathrm{d} {N_{i/0}^+}}{\mathrm{d} \varepsilon} + \eta_i \right] \Biggl]
    \\ & 
    \Biggl[ 1 - \frac{ f(\varepsilon) }{
    \left[ f(\varepsilon)^2 - 4 \sigma_c \sigma_g \rho \gamma^{-1} \langle v_{N} \rangle [\mu_I + \sigma_I z_{I, 0}] \right]^{\frac{1}{2}}}
    \Biggl].
\end{split}
\label{eq:dRde final}
\end{align}

The second moment of the sensitivity to perturbations, $\langle [d R_0^+ / d \varepsilon]^2 \rangle$, is 

\begin{align}
\begin{split}
    \left \langle \left[ \frac{ \mathrm{d} R_0}{\mathrm{d} \varepsilon} \right]^2 \right \rangle =&
    \frac{1}{4 [\sigma_c \sigma_g \rho \gamma^{-1} \langle v_{N} \rangle]^2} \Biggl \langle \Biggl[
    \frac{\sigma_c}{\sqrt{M}} \sum_{i=1}^{S^*} \left[ \rho z_{g, i 0} + \sqrt{1 - \rho^2} z_{c, i 0} \right] \left[ \frac{ \mathrm{d} {N_{i/0}^+}}{\mathrm{d} \varepsilon} + \eta_i \right] \Biggl]^2
    \\ & \hphantom{\frac{1}{4 [\sigma_c \sigma_g \rho \gamma^{-1} \langle v_{N} \rangle]^2} \Biggl \langle \Biggl[}
    \Biggl[ 
    1 
    - \frac{ 2 f(\varepsilon) }{ \left[ f(\varepsilon)^2 - 4 \sigma_c \sigma_g \rho \gamma^{-1} \langle v_{N} \rangle [\mu_I + \sigma_I z_{I, 0}] \right]^{\frac{1}{2}}}
    \\ & \hphantom{\frac{1}{4 [\sigma_c \sigma_g \rho \gamma^{-1} \langle v_{N} \rangle]^2} \Biggl \langle \Biggl[ \Biggl[]}
    + \frac{ f(\varepsilon)^2 }{f(\varepsilon)^2 - 4 \sigma_c \sigma_g \rho \gamma^{-1} \langle v_{N} \rangle [\mu_I + \sigma_I z_{I, 0}]}
    \Biggl] \Biggl \rangle
    \\ \approx &
    \frac{1}{4 \left[ \sigma_c \sigma_g \rho \gamma^{-1} \langle v_{N} \rangle \right]^2}
    \left \langle \Biggl[ \frac{\sigma_c}{\sqrt{M}} \sum_{i=1}^{S^*} \left[ \rho z_{g, i 0} + \sqrt{1 - \rho^2} z_{c, i 0} \right] \left[ \frac{ \mathrm{d} {N_{i/0}^+}}{\mathrm{d} \varepsilon} + \eta_i \right] \Biggl]^2 \right \rangle
    \\ & \hphantom{\frac{1}{4 \left[ \sigma_c \sigma_g \rho \gamma^{-1} \langle v_{N} \rangle \right]^2}}
    \Biggl \langle \Biggl[ 
    1 
    - \frac{ 2 f(\varepsilon) }{ \left[ f(\varepsilon)^2 - 4 \sigma_c \sigma_g \rho \gamma^{-1} \langle v_{N} \rangle [\mu_I + \sigma_I z_{I, 0}] \right]^{\frac{1}{2}}}
    \\ & \hphantom{\frac{1}{4 \left[ \sigma_c \sigma_g \rho \gamma^{-1} \langle v_{N} \rangle \right]^2} \Biggl \langle \Biggl[}
    + \frac{ f(\varepsilon)^2 }{f(\varepsilon)^2 - 4 \sigma_c \sigma_g \rho \gamma^{-1} \langle v_{N} \rangle [\mu_I + \sigma_I z_{I, 0}]}
    \Biggl] \Biggl \rangle.
\end{split}
\label{eq:dRde squared}
\end{align}

We can treat the two terms in the angle brackets as approximately independent because their correlation is subleading in the large-$M$ limit.
The expectation of the first term is

\begin{align}
\begin{split}
&
    \left \langle \Biggl[ \frac{\sigma_c}{\sqrt{M}} \sum_{i=1}^{S^*} \left[ \rho z_{g, i 0} + \sqrt{1 - \rho^2} z_{c, i 0} \right] \left[ \frac{ \mathrm{d} {N_{i/0}^+}}{\mathrm{d} \varepsilon} + \eta_i \right] \Biggl]^2 \right \rangle
    = \sigma_c^2 \phi_N \gamma^{-1} \left[ \left \langle \left[ \frac{ \mathrm{d} {N_{i/0}^+}}{\mathrm{d} \varepsilon} \right]^2 \right \rangle + 1 \right].
\end{split}
\label{eq:dRde squared term 1}
\end{align}

Turning to the second term in eq. \eqref{eq:dRde squared}, we cannot compute it in its current form, as it is a non-linear function of $f(\varepsilon)$.
To linearise the expression in terms of $f(\varepsilon)$, we note that $f(\varepsilon)^2 - 4 \sigma_c \sigma_g \rho \gamma^{-1} \langle v_{N} \rangle [\mu_I + \sigma_I z_{I, 0}]$ is composed of means and fluctuations:

\begin{align}
\begin{split}
&
    \left[  
        \mu_b + \mu_c \gamma^{-1} \langle N \rangle 
        + \frac{\sigma_c}{\sqrt{M}} \sum_{i=1}^{S^*} \left[ \rho z_{g, i 0} + \sqrt{1 - \rho^2} z_{c, i 0} \right] \left[ {N_{i/0}^+} +  \varepsilon \eta_i \right] + \sigma_b z_{b, 0} 
        \right]^2
        - 4 \sigma_c \sigma_g \rho \gamma^{-1} \langle v_{N} \rangle [\mu_I + \sigma_I z_{I, 0}]
\\ & =
\underbrace{ 
    \left[ \mu_b + \mu_c \gamma^{-1} \langle N \rangle \right]^2 - 4 \sigma_c \sigma_g \rho \gamma^{-1} \langle v_{N} \rangle \mu_I
}_{\text{means}}
\\ & \hphantom{=} + 
    2 \left[ \mu_b + \mu_c \gamma^{-1} \langle N \rangle \right] \left[ 
        \frac{\sigma_c}{\sqrt{M}} \sum_{i=1}^{S^*} \left[ \rho z_{g, i 0} + \sqrt{1 - \rho^2} z_{c, i 0} \right] \left[ {N_{i/0}^+} +  \varepsilon \eta_i \right] + \sigma_b z_{b, 0} 
    \right]
\\ & \hphantom{=} 
\underbrace{+  
\left[ 
        \frac{\sigma_c}{\sqrt{M}} \sum_{i=1}^{S^*} \left[ \rho z_{g, i 0} + \sqrt{1 - \rho^2} z_{c, i 0} \right] \left[ {N_{i/0}^+} +  \varepsilon \eta_i \right] + \sigma_b z_{b, 0} 
    \right]^2
- 4 \sigma_c \sigma_g \rho \gamma^{-1} \langle v_{N} \rangle \sigma_I z_{I, 0}
}_{\text{fluctuations}}.
\end{split}       
\label{eq:dRde squared term 2 denominator means fluctuations}
\end{align}

For shorthand, we will rewrite these as 

\begin{align} \begin{split}
    \underbrace{\mu(\varepsilon)^2 - 4 F \mu_I}_{\text{means}} 
    + \underbrace{ 2 \mu \sigma( \varepsilon )+ \sigma( \varepsilon )^2 - 4 F \sigma_I z_{I, 0}}_{\text{fluctuations}}.
\end{split} \end{align}

where $\mu(\varepsilon) = \mu_b + \mu_c \gamma^{-1} \langle N \rangle$, 
$F = \sigma_c \sigma_g \rho \gamma^{-1} \langle v_{N} \rangle$ 
and $\sigma(\varepsilon) = \frac{\sigma_c}{\sqrt{M}} \sum_{i=1}^{S^*} \left[ \rho z_{g, i 0} + \sqrt{1 - \rho^2} z_{c, i 0} \right] \left[ {N_{i/0}^+} +  \varepsilon \eta_i \right] + \sigma_b z_{b, 0}$.

Assuming that the means are far larger than the fluctuations, we can approximate this term by taking the linear binomial expansion about its mean.
Therefore,

\begin{align}
\begin{split}
    - \frac{ 2 f(\varepsilon) }{ \left[ f(\varepsilon)^2 - 4 \sigma_c \sigma_g \rho \gamma^{-1} \langle v_{N} \rangle [\mu_I + \sigma_I z_{I, 0}] \right]^{\frac{1}{2}}}
    \approx &
    - \frac{2 f(\varepsilon)}{\left[ \mu(\varepsilon)^2 - 4 F \mu_I \right]^{\frac{1}{2}}} \left[ 1 - \frac{2 \mu \sigma( \varepsilon )+ \sigma( \varepsilon )^2 - 4 F \sigma_I z_{I, 0}}{2 \left[ \mu(\varepsilon)^2 - 4 F \mu_I \right]} \right]
    \\ \approx &
    - \frac{2 \left[ \mu + \sigma(\varepsilon) \right]}{\left[ \mu(\varepsilon)^2 - 4 F \mu_I \right]^{\frac{1}{2}}} \left[ 1 - \frac{2 \mu \sigma( \varepsilon )+ \sigma( \varepsilon )^2 - 4 F \sigma_I z_{I, 0}}{2 \left[ \mu(\varepsilon)^2 - 4 F \mu_I \right]} \right]
\end{split}
\label{eq:dRde squared term 1 binom approx}
\end{align}

and 

\begin{align}
\begin{split}
    \frac{ f(\varepsilon)^2 }{f(\varepsilon)^2 - 4 \sigma_c \sigma_g \rho \gamma^{-1} \langle v_{N} \rangle [\mu_I + \sigma_I z_{I, 0}]}
    \approx &
    \frac{f(\varepsilon)^2}{\mu(\varepsilon)^2 - 4 F \mu_I} \left[ 1 - \frac{2 \mu \sigma( \varepsilon )+ \sigma( \varepsilon )^2 - 4 F \sigma_I z_{I, 0}}{\mu(\varepsilon)^2 - 4 F \mu_I} \right]
    \\ \approx &
    \frac{\left[ \mu + \sigma(\varepsilon) \right]^2}{\mu(\varepsilon)^2 - 4 F \mu_I} \left[ 1 - \frac{2 \mu \sigma( \varepsilon )+ \sigma( \varepsilon )^2 - 4 F \sigma_I z_{I, 0}}{\mu(\varepsilon)^2 - 4 F \mu_I} \right].
\end{split}
\label{eq:dRde squared term 2 binom approx}
\end{align}

Hence, the second term in angle brackets in eq. \eqref{eq:dRde squared} is approximately

\begin{align}
\begin{split}
&
    \left \langle 
    1 
    - \frac{2 \left[ \mu + \sigma(\varepsilon) \right]}{\left[ \mu(\varepsilon)^2 - 4 F \mu_I \right]^{\frac{1}{2}}} \left[ 1 - \frac{2 \mu \sigma( \varepsilon )+ \sigma( \varepsilon )^2 - 4 F \sigma_I z_{I, 0}}{2 \left[ \mu(\varepsilon)^2 - 4 F \mu_I \right]} \right]
    + \frac{\left[ \mu + \sigma(\varepsilon) \right]^2}{\mu(\varepsilon)^2 - 4 F \mu_I} \left[ 1 - \frac{2 \mu \sigma( \varepsilon )+ \sigma( \varepsilon )^2 - 4 F \sigma_I z_{I, 0}}{\mu(\varepsilon)^2 - 4 F \mu_I} \right]
    \right \rangle.
\\ & \approx
    1 
    - \frac{2}{\left[ \mu(\varepsilon)^2 - 4 F \mu_I \right]^{\frac{1}{2}}} \left[ 
      \mu - \frac{\mu \left[ 2 \mu \langle \sigma( \varepsilon ) \rangle + \langle \sigma( \varepsilon )^2 \rangle - 4 F \sigma_I \langle z_{I, 0} \rangle \right]}{2 \left[ \mu(\varepsilon)^2 - 4 F \mu_I \right]}
    +  \langle \sigma( \varepsilon ) \rangle - \frac{2 \mu \langle \sigma( \varepsilon )^2 \rangle + \langle \sigma( \varepsilon )^3 \rangle - 4 F \sigma_I \langle \sigma( \varepsilon ) \rangle \langle z_{I, 0} \rangle}{2 \left[ \mu(\varepsilon)^2 - 4 F \mu_I \right]} \right]
\\ &
    + \frac{1}{\mu(\varepsilon)^2 - 4 F \mu_I} 
        \Biggl[ \mu^2 - \frac{\mu^2 \left[ 2 \mu \langle \sigma( \varepsilon ) \rangle + \langle \sigma( \varepsilon )^2 \rangle - 4 F \sigma_I \langle z_{I, 0} \rangle \right]}{\mu(\varepsilon)^2 - 4 F \mu_I}
        + 2 \mu \langle \sigma( \varepsilon ) \rangle
\\ & \hphantom{+ \frac{1}{\mu(\varepsilon)^2 - 4 F \mu_I} \Biggl[}
         - \frac{ 2 \mu \left[ 2 \mu \langle \sigma( \varepsilon )^2 \rangle + \langle \sigma( \varepsilon )^3 \rangle - 4 F \sigma_I \langle \sigma( \varepsilon ) \rangle \langle z_{I, 0} \rangle \right]}{\mu(\varepsilon)^2 - 4 F \mu_I}
        + \langle \sigma( \varepsilon )^2 \rangle - \frac{2 \mu \langle \sigma( \varepsilon )^3 \rangle + \langle \sigma( \varepsilon )^4 \rangle - 4 F \sigma_I \langle \sigma( \varepsilon )^2 \rangle \langle z_{I, 0} \rangle}{\mu(\varepsilon)^2 - 4 F \mu_I} \Biggl].
\end{split}
\label{eq:dRde squared binom approx}
\end{align}

This expression has terms including the 1st to the 4th moment of $\sigma( \varepsilon )$.
For ease, we will calculate the non-zero components of these moments below, and later plug them into the terms in eq. \eqref{eq:dRde squared binom approx}:

\begin{align} \begin{split}
    \langle \sigma( \varepsilon ) \rangle
    = \frac{\sigma_c}{\sqrt{M}} \sum_{i=1}^{S^*} \left[ \rho \langle z_{g, i 0} \rangle + \sqrt{1 - \rho^2} \langle z_{c, i 0} \rangle \right] \left[ \langle {N_{i/0}^+} \rangle +  \varepsilon \langle \eta_i \rangle \right] + \sigma_b \langle z_{b, 0} \rangle 
    = 0.
\end{split} \end{align}

\begin{align} \begin{split}
    \langle \sigma( \varepsilon )^2 \rangle 
    &= \frac{\sigma_c^2}{M}
        \sum_i^{S^*} 
            \left[ \rho^2 \langle z_{g, i 0}^2 \rangle + 2 \rho \sqrt{1 - \rho^2} \langle z_{g, i 0} z_{c, i 0} \rangle  + [1 - \rho^2] \langle z_{c, i 0}^2 \rangle \right] 
            \left[ \langle {N_{i/0}^+}^2 \rangle +  2 \varepsilon \langle {N_{i/0}^+} \rangle \langle \eta_i \rangle + \varepsilon^2 \langle \eta_i^2 \rangle \right]
\\ & \hphantom{=}
    + \sigma_b^2\langle z_{b, 0}^2 \rangle
\\ & =
\sigma_c^2 \gamma^{-1} \phi_N \left[ \frac{\langle N^2 \rangle}{\phi_N} + \varepsilon^2 \right] + \sigma_b^2
\\ & \approx 
\sigma_c^2 \gamma^{-1} \langle N^2 \rangle + \sigma_b^2
\end{split} \end{align}

since $0 < \varepsilon \ll 1$.

\begin{align} \begin{split}
    \langle \sigma( \varepsilon )^3 \rangle = 0.
\end{split} \end{align}

{\allowdisplaybreaks
\begin{align} \begin{split}
    \langle \sigma( \varepsilon )^4 \rangle =&
    \frac{\sigma_c^4}{M^2} \Biggl[
    \sum_i^{S^*} 
        \Big[ 
            \rho^4 \langle z_{g, i 0}^4 \rangle 
            + 4 \rho^3 \sqrt{1 - \rho^2} \langle z_{g, i 0}^3 \rangle \langle z_{c, i 0} \rangle
            + 6 \rho^2 [1 - \rho^2] \langle z_{g, i 0}^2 \rangle \langle z_{c, i 0}^2 \rangle
\\ & \hphantom{\frac{\sigma_c^4}{M^2} \Biggl[ \sum_i^{S^*} \Big[}
            + 4 \rho [1 - \rho^2]^{\frac{3}{2}} \langle z_{g, i 0} \rangle \langle z_{c, i 0}^3 \rangle
            + [1 - \rho^2]^2 \langle z_{c, i 0}^4 \rangle \Big] 
\\ & \hphantom{\frac{\sigma_c^4}{M^2} \Biggl[ \sum_i^{S^*}}
        \Big[ 
            \langle {N_{i/0}^+}^4 \rangle 
            + 4 \varepsilon \langle {N_{i/0}^+}^3 \rangle \langle \eta_i \rangle 
            + 6 \varepsilon^2 \langle {N_{i/0}^+}^2 \rangle \langle \eta_i^2 \rangle
            + 4 \varepsilon^3 \langle {N_{i/0}^+} \rangle \langle \eta_i^3 \rangle
            + \varepsilon^4 \langle \eta_i^4 \rangle \Big]
\\ & \hphantom{\frac{\sigma_c^4}{M^2} \Biggl[}
    + 3 \sum_{i, j \neq i}^{S^*} 
        \left[ \rho^2 \langle z_{g, i 0}^2 \rangle + 2 \rho \sqrt{1 - \rho^2} \langle z_{g, i 0} z_{c, i 0} \rangle  + [1 - \rho^2] \langle z_{c, i 0}^2 \rangle \right] 
\\ & \hphantom{\frac{\sigma_c^4}{M^2} \Biggl[ + \sum_{i, j \neq i}^{S^*}}
        \left[ \rho^2 \langle z_{g, j 0}^2 \rangle + 2 \rho \sqrt{1 - \rho^2} \langle z_{g, j 0} z_{c, j 0} \rangle  + [1 - \rho^2] \langle z_{c, j 0}^2 \rangle \right] 
\\ & \hphantom{\frac{\sigma_c^4}{M^2} \Biggl[ + \sum_{i, j \neq i}^{S^*}}
        \left[ \langle {N_{i/0}^+}^2 \rangle +  2 \varepsilon \langle {N_{i/0}^+} \rangle \langle \eta_i \rangle + \varepsilon^2 \langle \eta_i^2 \rangle \right]
        \left[ \langle {N_{j/0}^+}^2 \rangle +  2 \varepsilon \langle {N_{j/0}^+} \rangle \langle \eta_j \rangle + \varepsilon^2 \langle \eta_j^2 \rangle \right]
    \Biggl]
\\ &
    + \frac{6 \sigma_c^2 \sigma_b^2}{M} \sum_i^{S^*} 
        \left[ \rho^2 \langle z_{g, i 0}^2 \rangle + 2 \rho \sqrt{1 - \rho^2} \langle z_{g, i 0} z_{c, i 0} \rangle  + [1 - \rho^2] \langle z_{c, i 0}^2 \rangle \right] 
 \\ & \hphantom{+ \frac{6 \sigma_c^2 \sigma_b^2}{M} \sum_i^{S^*}}       
        \left[ \langle {N_{i/0}^+}^2 \rangle +  2 \varepsilon \langle {N_{i/0}^+} \rangle \langle \eta_i \rangle + \varepsilon^2 \langle \eta_i^2 \rangle \right]
        \langle z_{I, 0}^2 \rangle
\\ &
    + \sigma_b^4 \langle z_{I, 0}^4 \rangle
\\ =&
    \frac{\sigma_c^4 \phi_N S }{M^2} 
        \left[ 3 \rho^4 + 6 \rho^2 - 6 \rho^4 + 3 - 6\rho^2 + 3\rho^4 \right]
        \left[ \langle {N_{i/0}^+}^4 \rangle + 6 \varepsilon^2 \frac{\langle N^2 \rangle}{\phi_N} + 3 \varepsilon^4 \right]
\\ &
    + \frac{\sigma_c^4 \phi_N^2 S^2 }{M^2} 
        \left[ \rho^2 + 1 - \rho^2 \right]^2
        \left[ \frac{\langle N^2 \rangle}{\phi_N} + \varepsilon^2 \right]^2
    + \frac{6 \sigma_c^2 \sigma_b^2 \phi_N S}{M} 
        \left[ \rho^2 + 1 - \rho^2 \right]
        \left[ \frac{\langle N^2 \rangle}{\phi_N} + \varepsilon^2 \right]
    + 3 \sigma_b^4
\\ \approx &
    3 \left[ \sigma_c^2 \phi_N \gamma^{-1} \right]^2 \left[ \frac{\langle N^2 \rangle}{\phi_N} \right]^2
    + 6 \sigma_c^2 \sigma_b^2 \phi_N \gamma^{-1} \left[ \frac{\langle N^2 \rangle}{\phi_N} \right]
    + 3 \sigma_b^4
\\ \approx &
    3 \left[ \sigma_c^2 \gamma^{-1} \langle N^2 \rangle \right]^2
    + 6 \sigma_b^2 \left[ \sigma_c^2 \gamma^{-1} \langle N^2 \rangle \right] 
    + 3 \sigma_b^4
\\ \approx &
    3 \left[ \sigma_c^2 \gamma^{-1} \langle N^2 \rangle + \sigma_b^2 \right]^2.
\end{split} \end{align}
}

as $0 < \varepsilon \ll 1$ and $M \rightarrow \infty$.

Returning to eq. \eqref{eq:dRde squared binom approx} plugging in our moments of $\sigma(\varepsilon)$, as well as $\mu(\varepsilon)$ and $F$ reduces the expression to 

\begin{align}
\begin{split}
    &
    1 
    - \frac{2}{\left[ \mu(\varepsilon)^2 - 4 F \mu_I \right]^{\frac{1}{2}}} \left[ 
        \mu 
        - \frac{3 \mu \langle \sigma( \varepsilon )^2 \rangle}{2 \left[ \mu(\varepsilon)^2 - 4 F \mu_I \right]} \right]
    + \frac{1}{\mu(\varepsilon)^2 - 4 F \mu_I} \Biggl[ 
        \mu^2 + \langle \sigma( \varepsilon )^2 \rangle 
        - \frac{5 \mu^2 \langle \sigma( \varepsilon )^2 \rangle + \langle \sigma( \varepsilon )^4 \rangle}{\mu(\varepsilon)^2 - 4 F \mu_I} \Biggl]
    \\ & \approx
    1 
    - \frac{2}{\left[ \left[ \mu_b + \mu_c \gamma^{-1} \langle N \rangle \right]^2 - 4 \sigma_c \sigma_g \rho \gamma^{-1} \langle v_{N} \rangle \mu_I \right]^{\frac{1}{2}}} \left[ 
        \mu_b + \mu_c \gamma^{-1} \langle N \rangle
        - \frac{3 \left[ \mu_b + \mu_c \gamma^{-1} \langle N \rangle\right] \left[ \sigma_c^2 \gamma^{-1} \langle N^2 \rangle + \sigma_b^2 \right] }{2 \left[ \left[ \mu_b + \mu_c \gamma^{-1} \langle N \rangle \right]^2 - 4 \sigma_c \sigma_g \rho \gamma^{-1} \langle v_{N} \rangle \mu_I \right]} \right]
     \\ & \hphantom{\approx}
     + \frac{1}{\left[ \mu_b + \mu_c \gamma^{-1} \langle N \rangle \right]^2 - 4 \sigma_c \sigma_g \rho \gamma^{-1} \langle v_{N} \rangle \mu_I} \Biggl[ 
        \left[ \mu_b + \mu_c \gamma^{-1} \langle N \rangle \right]^2 +\sigma_c^2 \gamma^{-1} \langle N^2 \rangle + \sigma_b^2 
    \\ & \hphantom{\approx + \frac{1}{\left[ \mu_b + \mu_c \gamma^{-1} \langle N \rangle \right]^2 - 4 \sigma_c \sigma_g \rho \gamma^{-1} \langle v_{N} \rangle \mu_I} \Biggl[}
        - \frac{5 \left[ \mu_b + \mu_c \gamma^{-1} \langle N \rangle \right]^2 \left[ \sigma_c^2 \gamma^{-1} \langle N^2 \rangle + \sigma_b^2 \right]}{\left[ \mu_b + \mu_c \gamma^{-1} \langle N \rangle \right]^2 - 4 \sigma_c \sigma_g \rho \gamma^{-1} \langle v_{N} \rangle \mu_I}
    \\ & \hphantom{\approx + \frac{1}{\left[ \mu_b + \mu_c \gamma^{-1} \langle N \rangle \right]^2 - 4 \sigma_c \sigma_g \rho \gamma^{-1} \langle v_{N} \rangle \mu_I} \Biggl[}    
    - \frac{
    3 \left[ \sigma_c^2 \gamma^{-1} \langle N^2 \rangle
    + \sigma_b^2 \right]^2}{\left[ \mu_b + \mu_c \gamma^{-1} \langle N \rangle \right]^2 - 4 \sigma_c \sigma_g \rho \gamma^{-1} \langle v_{N} \rangle \mu_I} \Biggl].
\end{split}
\end{align}

Therefore, the final approximation for $\langle [ \mathrm{d} R_0 / \mathrm{d} \varepsilon ]^2 \rangle$ is 

\begin{align}
\begin{split}
    \left \langle \left[ \frac{ \mathrm{d} R_0}{\mathrm{d} \varepsilon} \right]^2 \right \rangle \approx &
    \frac{\sigma_c^2 \phi_N \gamma^{-1}}{4 \left[ \sigma_c \sigma_g \rho \gamma^{-1} \langle v_{N} \rangle \right]^2}
     \left[ \left \langle \left[ \frac{ \mathrm{d} {N_{i/0}^+}}{\mathrm{d} \varepsilon} \right]^2 \right \rangle + 1 \right]
    \\ & \times
    \Biggl[ 1 
    - \frac{2}{\left[ \left[ \mu_b + \mu_c \gamma^{-1} \langle N \rangle \right]^2 - 4 \sigma_c \sigma_g \rho \gamma^{-1} \langle v_{N} \rangle \mu_I \right]^{\frac{1}{2}}} \Biggl[ 
        \mu_b + \mu_c \gamma^{-1} \langle N \rangle
    \\ & \hphantom{\times \Biggl[ 1 - \frac{2}{\left[ \left[ \mu_b + \mu_c \gamma^{-1} \langle N \rangle \right]^2 - 4 \sigma_c \sigma_g \rho \gamma^{-1} \langle v_{N} \rangle \mu_I \right]^{\frac{1}{2}}} \Biggl[} 
        - \frac{3 \left[ \mu_b + \mu_c \gamma^{-1} \langle N \rangle\right] \left[ \sigma_c^2 \gamma^{-1} \langle N^2 \rangle + \sigma_b^2 \right] }{2 \left[ \left[ \mu_b + \mu_c \gamma^{-1} \langle N \rangle \right]^2 - 4 \sigma_c \sigma_g \rho \gamma^{-1} \langle v_{N} \rangle \mu_I \right]} \Biggl]
     \\ & \hphantom{\times \Biggl[}
     + \frac{1}{\left[ \mu_b + \mu_c \gamma^{-1} \langle N \rangle \right]^2 - 4 \sigma_c \sigma_g \rho \gamma^{-1} \langle v_{N} \rangle \mu_I} \Biggl[ 
        \left[ \mu_b + \mu_c \gamma^{-1} \langle N \rangle \right]^2 +\sigma_c^2 \gamma^{-1} \langle N^2 \rangle + \sigma_b^2 
    \\ & \hphantom{\times \Biggl[ + \frac{1}{\left[ \mu_b + \mu_c \gamma^{-1} \langle N \rangle \right]^2 - 4 \sigma_c \sigma_g \rho \gamma^{-1} \langle v_{N} \rangle \mu_I} \Biggl[}
        - \frac{5 \left[ \mu_b + \mu_c \gamma^{-1} \langle N \rangle \right]^2 \left[ \sigma_c^2 \gamma^{-1} \langle N^2 \rangle + \sigma_b^2 \right]}{\left[ \mu_b + \mu_c \gamma^{-1} \langle N \rangle \right]^2 - 4 \sigma_c \sigma_g \rho \gamma^{-1} \langle v_{N} \rangle \mu_I}
    \\ & \hphantom{\times \Biggl[ + \frac{1}{\left[ \mu_b + \mu_c \gamma^{-1} \langle N \rangle \right]^2 - 4 \sigma_c \sigma_g \rho \gamma^{-1} \langle v_{N} \rangle \mu_I} \Biggl[}    
    - \frac{
    3 \left[ \sigma_c^2 \gamma^{-1} \langle N^2 \rangle 
    + \sigma_b^2 \right]^2}{\left[ \mu_b + \mu_c \gamma^{-1} \langle N \rangle \right]^2 - 4 \sigma_c \sigma_g \rho \gamma^{-1} \langle v_{N} \rangle \mu_I} \Biggl]
    \Biggl].
\end{split}
\end{align}

Because the cavity consumer is statistically identical to all community members, we can substitute $ \langle [ \mathrm{d} N_{i \wo 0}^+ / \mathrm{d} \varepsilon ]^2 \rangle$ with $ \langle [ \mathrm{d} N_0 / \mathrm{d} \varepsilon ]^2 \rangle$:

\begin{align}
\begin{split}
    \left \langle \left[ \frac{ \mathrm{d} R_0}{\mathrm{d} \varepsilon} \right]^2 \right \rangle \approx &
    \frac{\phi_N \gamma^{-1}}{4 \left[ \sigma_g \rho \gamma^{-1} \langle v_{N} \rangle \right]^2}
     \left[ \left \langle \left[ \frac{ \mathrm{d} {N_0^+}}{\mathrm{d} \varepsilon} \right]^2 \right \rangle + 1 \right]
    \\ & \times
    \Biggl[ 1 
    - \frac{2}{\left[ \left[ \mu_b + \mu_c \gamma^{-1} \langle N \rangle \right]^2 - 4 \sigma_c \sigma_g \rho \gamma^{-1} \langle v_{N} \rangle \mu_I \right]^{\frac{1}{2}}} \Biggl[ 
        \mu_b + \mu_c \gamma^{-1} \langle N \rangle
    \\ & \hphantom{\times \Biggl[ 1 - \frac{2}{\left[ \left[ \mu_b + \mu_c \gamma^{-1} \langle N \rangle \right]^2 - 4 \sigma_c \sigma_g \rho \gamma^{-1} \langle v_{N} \rangle \mu_I \right]^{\frac{1}{2}}} \Biggl[} 
        - \frac{3 \left[ \mu_b + \mu_c \gamma^{-1} \langle N \rangle\right] \left[ \sigma_c^2 \gamma^{-1} \langle N^2 \rangle + \sigma_b^2 \right] }{2 \left[ \left[ \mu_b + \mu_c \gamma^{-1} \langle N \rangle \right]^2 - 4 \sigma_c \sigma_g \rho \gamma^{-1} \langle v_{N} \rangle \mu_I \right]} \Biggl]
     \\ & \hphantom{\times \Biggl[}
     + \frac{1}{\left[ \mu_b + \mu_c \gamma^{-1} \langle N \rangle \right]^2 - 4 \sigma_c \sigma_g \rho \gamma^{-1} \langle v_{N} \rangle \mu_I} \Biggl[ 
        \left[ \mu_b + \mu_c \gamma^{-1} \langle N \rangle \right]^2 +\sigma_c^2 \gamma^{-1} \langle N^2 \rangle + \sigma_b^2 
    \\ & \hphantom{\times \Biggl[ + \frac{1}{\left[ \mu_b + \mu_c \gamma^{-1} \langle N \rangle \right]^2 - 4 \sigma_c \sigma_g \rho \gamma^{-1} \langle v_{N} \rangle \mu_I} \Biggl[}
        - \frac{5 \left[ \mu_b + \mu_c \gamma^{-1} \langle N \rangle \right]^2 \left[ \sigma_c^2 \gamma^{-1} \langle N^2 \rangle + \sigma_b^2 \right]}{\left[ \mu_b + \mu_c \gamma^{-1} \langle N \rangle \right]^2 - 4 \sigma_c \sigma_g \rho \gamma^{-1} \langle v_{N} \rangle \mu_I}
    \\ & \hphantom{\times \Biggl[ + \frac{1}{\left[ \mu_b + \mu_c \gamma^{-1} \langle N \rangle \right]^2 - 4 \sigma_c \sigma_g \rho \gamma^{-1} \langle v_{N} \rangle \mu_I} \Biggl[}    
    - \frac{
    3 \left[ \sigma_c^2 \gamma^{-1} \langle N^2 \rangle 
    + \sigma_b^2 \right]^2}{\left[ \mu_b + \mu_c \gamma^{-1} \langle N \rangle \right]^2 - 4 \sigma_c \sigma_g \rho \gamma^{-1} \langle v_{N} \rangle \mu_I} \Biggl]
    \Biggl].
\end{split}
\end{align}

\subsubsection{Stability condition}

Solving for $\langle \left[ \mathrm{d}N_0^+ / d \varepsilon \right]^2 \rangle$ and $\langle \left[ \mathrm{d}R_0 / \mathrm{d} \varepsilon \right]^2 \rangle$ gives us expressions of the form 
$ \langle [ \mathrm{d}N_0^+ / d \varepsilon ]^2 \rangle = [B+1] / [AB-1]$ and 
$\langle [ \mathrm{d}R_0 / \mathrm{d} \varepsilon ]^2 \rangle = [A+1]/[AB-1]$, 
where $1/A = 1 / [ \sigma_c \rho \langle \chi_{R} \rangle ]^2$ and 
$1/B = [\phi_N \gamma^{-1} / 4 [\sigma_g \gamma^{-1} \rho \langle v_{N} \rangle]^2] \times 
 [\text{the long expression in } \langle [ \mathrm{d}R_0 / \mathrm{d} \varepsilon ]^2 \rangle]$
Communities undergo a transition to instability when these sensitivities diverge, which occurs when $AB-1=0$.
Therefore, communities are stable when $AB-1 > 0$ or equivalently when $1/AB < 1$:

\begin{align}
\begin{split}
    \frac{\phi_N \gamma^{-1}}{4[\sigma_c \rho \langle \chi_{R} \rangle]^2[\sigma_g \gamma^{-1} \rho \langle v_{N} \rangle]^2}
    \Biggl[ ... \Biggl]
    < 1.
\end{split}
\end{align}

We can simplify the expression outside the bracket by plugging in the expression for $\langle v_{N} \rangle$ in eq. \eqref{eq:v_N}, which reduces the expression to

\begin{align}
\begin{split}
    \frac{1}{4 \rho^2 \phi_N \gamma^{-1}} \Biggl[... \Biggl] < 1,
    \text{ or equivalently, }
    \frac{1}{4 \rho^2} \Biggl[ ... \Biggl] < \phi_N \gamma^{-1}.
    \label{eq:stab_cond_outside_bracket}
\end{split}
\end{align}

Therefore, the full stability condition is 

\begin{tcolorbox}[colback=white!5!white,colframe=black!100!black,
title = Stability condition for the externally-supplied model,
fonttitle = \Large\bfseries]

\begin{align}
\begin{split}
&
    \frac{1}{4 \rho^2} \Biggl[
    1 
    - \frac{2}{\left[ \left[ \mu_b + \mu_c \gamma^{-1} \langle N \rangle \right]^2 - 4 \sigma_c \sigma_g \rho \gamma^{-1} \langle v_{N} \rangle \mu_I \right]^{\frac{1}{2}}} \Biggl[ 
        \mu_b + \mu_c \gamma^{-1} \langle N \rangle
    \\ & \hphantom{\frac{1}{4 \rho^2} \Biggl[ 1 - \frac{2}{\left[ \left[ \mu_b + \mu_c \gamma^{-1} \langle N \rangle \right]^2 - 4 \sigma_c \sigma_g \rho \gamma^{-1} \langle v_{N} \rangle \mu_I \right]^{\frac{1}{2}}} \Biggl[} 
        - \frac{3 \left[ \mu_b + \mu_c \gamma^{-1} \langle N \rangle\right] \left[ \sigma_c^2 \gamma^{-1} \langle N^2 \rangle + \sigma_b^2 \right] }{2 \left[ \left[ \mu_b + \mu_c \gamma^{-1} \langle N \rangle \right]^2 - 4 \sigma_c \sigma_g \rho \gamma^{-1} \langle v_{N} \rangle \mu_I \right]} \Biggl]
     \\ & \hphantom{\frac{1}{4 \rho^2} \Biggl[}
     + \frac{1}{\left[ \mu_b + \mu_c \gamma^{-1} \langle N \rangle \right]^2 - 4 \sigma_c \sigma_g \rho \gamma^{-1} \langle v_{N} \rangle \mu_I} \Biggl[ 
        \left[ \mu_b + \mu_c \gamma^{-1} \langle N \rangle \right]^2 +\sigma_c^2 \gamma^{-1} \langle N^2 \rangle + \sigma_b^2 
    \\ & \hphantom{\frac{1}{4 \rho^2} \Biggl[ + \frac{1}{\left[ \mu_b + \mu_c \gamma^{-1} \langle N \rangle \right]^2 - 4 \sigma_c \sigma_g \rho \gamma^{-1} \langle v_{N} \rangle \mu_I} \Biggl[}
        - \frac{5 \left[ \mu_b + \mu_c \gamma^{-1} \langle N \rangle \right]^2 \left[ \sigma_c^2 \gamma^{-1} \langle N^2 \rangle + \sigma_b^2 \right]}{\left[ \mu_b + \mu_c \gamma^{-1} \langle N \rangle \right]^2 - 4 \sigma_c \sigma_g \rho \gamma^{-1} \langle v_{N} \rangle \mu_I}
    \\ & \hphantom{\frac{1}{4 \rho^2} \Biggl[ + \frac{1}{\left[ \mu_b + \mu_c \gamma^{-1} \langle N \rangle \right]^2 - 4 \sigma_c \sigma_g \rho \gamma^{-1} \langle v_{N} \rangle \mu_I} \Biggl[}    
    - \frac{
    3 \left[ \sigma_c^2 \gamma^{-1} \langle N^2 \rangle
    + \sigma_b^2 \right]^2}{\left[ \mu_b + \mu_c \gamma^{-1} \langle N \rangle \right]^2 - 4 \sigma_c \sigma_g \rho \gamma^{-1} \langle v_{N} \rangle \mu_I} \Biggl]
    \Biggl] 
\\ &
    < \phi_N \gamma^{-1}.
\end{split}
\label{eq:big_stability_condition}
\end{align}
\end{tcolorbox}

\subsection{Infeasibility condition}

Next, we will investigate whether and when community dynamics become infeasible i.e., when a solution for the cavity self-consistency equations do not exist.
From inspecting the self-consistency equations, we can see that the consumer and resource abundance distributions diverge to infinity when when $\langle \chi_{R} \rangle \rightarrow 0$ and $\langle v_{N} \rangle \rightarrow -\infty$.
To determine when the system tends towards these limits, we will partially solve for $\langle v_{N} \rangle$ and $\langle \chi_{R} \rangle$.
Plugging in eq. \eqref{eq:v_N} to eq. \eqref{eq:chi_R} and rearranging yields the following expression:

\begin{align}
\begin{split}
    \phi_N \gamma^{-1}= 
    \frac{1}{2}
    \Biggl[
    1 
    - \frac{\mu_b + \mu_c \gamma^{-1} \langle N \rangle}{\sqrt{[\mu_b + \mu_c \gamma^{-1} \langle N \rangle]^2 
            - 4 \sigma_c \sigma_g \rho \gamma^{-1} \langle v_{N} \rangle \mu_I}}
    \Bigg[ 
    1 - 
    \frac{3 \left[ \sigma_c^2 \gamma^{-1} \langle N^2 \rangle + \sigma_b^2 \right]}{
    2 \left[[\mu_b + \mu_c \gamma^{-1} \langle N \rangle]^2 
            - 4 \sigma_c \sigma_g \rho \gamma^{-1} \langle v_{N} \rangle \mu_I\right]}
    \Biggl] \Biggl].
\end{split}
\label{eq:chi_partial_solve}
\end{align}

Since $\langle v_{N} \rangle \to -\infty$, $\langle N \rangle \to \infty$ and $\langle N^2 \rangle \to \infty$ in the infeasible limit, one can make a heuristic argument that $[\mu_b + \mu_c \gamma^{-1} \langle N \rangle]  \div \sqrt{[\mu_b + \mu_c \gamma^{-1} \langle N \rangle]^2 - 4 \sigma_c \sigma_g \rho \gamma^{-1} \langle v_{N} \rangle \mu_I} \to 0$ since the denominator contains two infinities whereas the numerator contains 1.
Therefore, the r.h.s of eq. \eqref{eq:chi_partial_solve} tends to $1/2$ as the system tends towards infeasibility. 
Therefore, $\phi_N \gamma^{-1}$ i.e., the consumer packing ratio can never exceed $1/2$, consistent with ~\cite{cui_diverse_2021}.

\section{Unified stability condition}

We have thus far seen that the stability condition for the externally-supplied model, as well as the hybrid model, look rather different to the that of the self-renewing model~\cite{blumenthal_phase_2024,liu_complex_2025,liu_ecosystem_2024}. 
To better understand the relationship between them, we now present an alternate derivation of the stability condition which will turn out to crucially rely on how resources respond to perturbations in their own depletion rates, i.e., on the statistics of the resource self-susceptibility $\chi_{00}$. 
We will also show that this alternate derivation leads to a unified stability condition that reduces to the known stability conditions for both models in some simple limits. 
Towards the end of this section, we also use the ideas developed here to discuss stability conditions for other classes of models with different resource dynamics, as well as reconcile apparent inconsistencies in the literature.

\subsection{Sensitivity of resource abundances to perturbation}

As above, we will perturb the surviving consumers as $N_i^+\rightarrow N_i^++\varepsilon\eta_i$ and the resources as $R_\alpha\rightarrow R_\alpha+\varepsilon\eta_\alpha^{(R)}$. 
What distinguishes different models are their resource dynamics, and thus their resource sensitivities. 
We will first derive the stability condition for the externally-supplied resource model, paying close attention to the resource self-susceptibilities.
As in the preceding calculation, for ease of notation, we will first collect the total resource depletion, both by dilution and by consumers, into the field $f$ as

\begin{align}
\begin{split}
    f
    ={}&
    b_0 + \sum_i^S c_{i0} N_{i \wo 0}^+
    \\
    ={}&\mu_b+\mu_c\gamma^{-1}\langle N\rangle
    +\frac{\sigma_c}{\sqrt M}\sum_{i=1}^{S^*}
    \left[
    \rho z_{g,i0}+\sqrt{1-\rho^2}\,z_{c,i0}
    \right]N_{i\wo0}^+
    +\sigma_bz_{b,0}.
\end{split}
\label{eq:alternative f definition}
\end{align}

Writing 
\begin{equation}
F=\sigma_c\sigma_g\rho\gamma^{-1}\langle v_{N} \rangle
\quad
{\rm and}
\quad 
I_0=\mu_I+\sigma_Iz_{I,0},
\label{eq:f_def}
\end{equation}
the cavity resource abundance may be written as the solution of a quadratic equation as
\begin{align}
    R_0=\frac{f-\sqrt{f^2-4FI_0}}{2F}.
\label{eq:alternative R0 compact}
\end{align}

The dilution rate $b_0$ and the depletion generated by consumers enter $R_0$ through the same field $f$.
By invoking the chain rule, we can characterize both responses using the diagonal resource susceptibility

\begin{align}
\begin{split}
    \chi_{00}
    &=-\frac{\partial R_0}{\partial b_0}
    =-\frac{\partial R_0}{\partial f} \frac{\partial f}{\partial b_0}
    =-\frac{\partial R_0}{\partial f}
    \\ &=-\frac{1}{2F}
    \left[
    1-\frac{f}{\sqrt{f^2-4FI_0}}
    \right]
    =\frac{R_0}{\sqrt{f^2-4FI_0}}.
\end{split}
\label{eq:chi00 closed}
\end{align}

Note that this expression does not involve the binomial expansion and approximation used in the previous calculation.
Averaging over the cavity ensemble results in the mean self-susceptibility we already used in the self-consistency equations previously, $\langle \chi_{R} \rangle=\langle\chi_{00}\rangle$.

We will now perturb the abundances of the surviving consumer species.
This changes the local field $f$ and hence we can write

\begin{align}
\begin{split}
    \frac{\mathrm{d}f}{\mathrm{d}\varepsilon}
    ={}&\frac{\sigma_c}{\sqrt M}
    \sum_{i=1}^{S^*}
    \left[
    \rho z_{g,i0}+\sqrt{1-\rho^2}\,z_{c,i0}
    \right]
    \left[
    \frac{\mathrm{d}N_{i\wo0}^+}{\mathrm{d}\varepsilon}
    +\eta_i
    \right].
\end{split}
\label{eq:alternative dfde}
\end{align}

Since the consumers affect $R_0$ only through $f$, we can use the chain rule to get an expression for how the resource abundance responds to perturbations in the consumer abundances

\begin{align}
    \frac{\mathrm{d}R_0}{\mathrm{d}\varepsilon}
    = \frac{\mathrm{d}R_0}{\mathrm{d}f} \frac{\mathrm{d}f}{\mathrm{d}\varepsilon}
    =-\chi_{00}\frac{\mathrm{d}f}{\mathrm{d}\varepsilon}.
\label{eq:alternative dRde}
\end{align}

Note again that Eqs.~\eqref{eq:chi00 closed} and \eqref{eq:alternative dRde} do not involve the binomial expansion and approximation.
To obtain the mean squared response, we now square Eq.~\eqref{eq:alternative dRde} and average over cavity realizations.
We will assume that in the large-$M$ limit, these two terms are weakly correlated, as in the previous calculation, leading to

\begin{align}
\begin{split}
    \left\langle
    \left[\frac{\mathrm{d}R_0}{\mathrm{d}\varepsilon}\right]^2
    \right\rangle
    &=\left\langle
    \chi_{00}^2
    \left[\frac{\mathrm{d}f}{\mathrm{d}\varepsilon}\right]^2
    \right\rangle
    \approx
    \langle\chi_{00}^2\rangle
    \left\langle
    \left[\frac{\mathrm{d}f}{\mathrm{d}\varepsilon}\right]^2
    \right\rangle.
\end{split}
\label{eq:alternative resource factorization}
\end{align}

The first of these factors is the second moment of the resource self-susceptibility and will be a key quantity in the unified stability condition; we will come to it later. 
We next evaluate the second factor by squaring Eq.~\eqref{eq:alternative dfde}.
This produces a double sum over pairs of surviving consumers:

\begin{align}
\begin{split}
    \left\langle
    \left[\frac{\mathrm{d}f}{\mathrm{d}\varepsilon}\right]^2
    \right\rangle
    ={}&\frac{\sigma_c^2}{M}
    \sum_{i=1}^{S^*}\sum_{j=1}^{S^*}
    \Biggl\langle
    \left[
    \rho z_{g,i0}+\sqrt{1-\rho^2}\,z_{c,i0}
    \right]
    \left[
    \rho z_{g,j0}+\sqrt{1-\rho^2}\,z_{c,j0}
    \right]
    \left[
    \frac{\mathrm{d}N_{i\wo0}^+}{\mathrm{d}\varepsilon}
    +\eta_i
    \right]
    \left[
    \frac{\mathrm{d}N_{j\wo0}^+}{\mathrm{d}\varepsilon}
    +\eta_j^{(N)}
    \right]
    \Biggr\rangle,
\end{split}
\label{eq:alternative dfde double sum}
\end{align}

which was calculated in eq. \eqref{eq:dRde squared term 1} to be approximately

\begin{align}
\begin{split}
    \left\langle
    \left[\frac{\mathrm{d}f}{\mathrm{d}\varepsilon}\right]^2
    \right\rangle
    \approx{}&
    \sigma_c^2\phi_N\gamma^{-1}
    \left[
    \left\langle
    \left[\frac{\mathrm{d}N_0^+}{\mathrm{d}\varepsilon}\right]^2
    \right\rangle+1
    \right].
\end{split}
\label{eq:alternative perturbation variance}
\end{align}

Substituting this result into eq.~\eqref{eq:alternative resource factorization} gives

\begin{align}
    \left\langle
    \left[\frac{\mathrm{d}R_0}{\mathrm{d}\varepsilon}\right]^2
    \right\rangle
    \approx\sigma_c^2\phi_N\gamma^{-1}
    \langle\chi_{00}^2\rangle
    \left[
    \left\langle
    \left[\frac{\mathrm{d}N_0^+}{\mathrm{d}\varepsilon}\right]^2
    \right\rangle+1
    \right].
\label{eq:resource half penultimate}
\end{align}

Since all resources and consumers are statistically identical, $\langle \chi_{R, 00}^2 \rangle = \langle \chi_{R, \alpha \alpha}^2 \rangle = \langle \chi_{R}^2 \rangle$, meaning eq. \eqref{eq:resource half penultimate} can be rewritten as 

\begin{align}
    \left\langle
    \left[\frac{\mathrm{d}R_0}{\mathrm{d}\varepsilon}\right]^2
    \right\rangle
    \approx\sigma_c^2\phi_N\gamma^{-1}
    \langle\chi_{R}^2\rangle
    \left[
    \left\langle
    \left[\frac{\mathrm{d}N_0^+}{\mathrm{d}\varepsilon}\right]^2
    \right\rangle+1
    \right].
\label{eq:resource half final}
\end{align}

This is an alternate compact form of the second moment of the resource response to perturbations in consumer abundances, analogous to Eq.~\eqref{eq:dRde final}.
The factor $\phi_N\gamma^{-1}=S^*/M$ appears because only the $S^*$ surviving consumers contribute to the response of a resource.
For another form of the resource supply dynamics, the expression for the resource self-susceptibility $\chi_{00}$ and its associated second moment $\langle\chi_{R}^2\rangle$ will change, but the form of Eq.~\eqref{eq:resource half final} will typically remain the same whenever depletion by consumers enters through the same scalar field $f$ as resource dilution $b_0$. 
Note that for the self-renewing resource model, the depletion by consumers enters through changing the effective resource growth rate rather than the dilution rate, but mathematically, these conditions are equivalent since resource growth (or supply) rate in the self-renewing model enters as a linear term $\propto R_0$ in the resource supply dynamics~\cite{blumenthal_phase_2024}.

\subsection{Combined condition}

We now combine the resource result with the consumer sensitivity derived in eq.~\eqref{eq:dNde2_cavity}. 
It tells us how a change in resource abundance changes a consumer.
Equation~\eqref{eq:resource half final} tells us how the resulting change in consumer abundance acts back on a resource.
Stability is lost when these moments diverge and do not have a well-defined solution. 
Biologically, this corresponds to the case where perturbations in resources affect consumers, and those consumer changes feed back and generate a further change in resources.
When this repeated response feedback becomes greatly amplified, the second moments diverge and the community becomes unstable.

To see this directly, we substitute eq.~\eqref{eq:resource half final} into eq.~\eqref{eq:dNde2_cavity}:

\begin{align}
\begin{split}
    \left\langle
    \left[\frac{\mathrm{d}N_0^+}{\mathrm{d}\varepsilon}\right]^2
    \right\rangle
    \approx{}&
    \frac{1}{\sigma_c^2\rho^2 \langle \chi_{R} \rangle^2}
    \Biggl[
    \sigma_c^2\phi_N\gamma^{-1}
    \langle\chi_{R}^2\rangle
    \left[
    \left\langle
    \left[\frac{\mathrm{d}N_0^+}{\mathrm{d}\varepsilon}\right]^2
    \right\rangle+1
    \right]+1
    \Biggr].
\end{split}
\label{eq:alternative response substitution}
\end{align}

Collecting the terms that contain the consumer sensitivity gives

\begin{align}
\begin{split}
    \left[
    \rho^2 \langle \chi_{R} \rangle^2
    -\phi_N\gamma^{-1}\langle\chi_{R}^2\rangle
    \right]
    \left\langle
    \left[\frac{\mathrm{d}N_0^+}{\mathrm{d}\varepsilon}\right]^2
    \right\rangle
    \approx
    \phi_N\gamma^{-1}\langle\chi_{R}^2\rangle
    +\frac{1}{\sigma_c^2}.
\end{split}
\label{eq:alternative response collected}
\end{align}

Thus,

\begin{align}
\begin{split}
    \left\langle
    \left[\frac{\mathrm{d}N_0^+}{\mathrm{d}\varepsilon}\right]^2
    \right\rangle
    \approx
    \frac{
    \phi_N\gamma^{-1}\langle\chi_{R}^2\rangle
    +1/\sigma_c^2
    }{
    \rho^2 \langle \chi_{R} \rangle^2
    -\phi_N\gamma^{-1}\langle\chi_{R}^2\rangle
    }.
\end{split}
\label{eq:alternative consumer sensitivity solution}
\end{align}

The resource sensitivity in eq.~\eqref{eq:resource half final} diverges at the same point.
Both sensitivities therefore remain finite and the community remains stable only when the denominator in eq.~\eqref{eq:alternative consumer sensitivity solution} is positive:

\begin{align}
    \rho^2 \langle \chi_{R} \rangle^2
    >\phi_N\gamma^{-1}\langle\chi_{R}^2\rangle.
\label{eq:alternative stability inequality}
\end{align}

Since $\phi_N\gamma^{-1}=S^*/M$, the stability condition becomes

\begin{tcolorbox}[colback=white!5!white,colframe=black!100!black,
title = Unified stability condition,
fonttitle = \Large\bfseries]

\begin{align}
    \rho^2>
    \frac{S^*}{M}
    \frac{\langle\chi_{R}^2\rangle}{ \langle \chi_{R} \rangle^2}.
\label{eq:stability condition}
\end{align}
\end{tcolorbox}

Note that this boxed stability condition applies to both the externally-supplied and self-renewing resource models, due to arguments we have made previously.
This stability condition has a rather simple interpretation.
The packing fraction $S^*/M$ counts how many surviving consumers contribute to the response of each resource, and is a standard and known determinant of ecological stability.
The mean susceptibility $\langle \chi_{R} \rangle=\langle\chi_{00}\rangle$ enters when a resource perturbation changes the consumers, whereas the second moment $\langle\chi_{R}^2\rangle$ enters when the consumer response acts back on the resources.
Their normalized ratio therefore analytically describes how including explicit resource dynamics changes the stability condition for complex ecological communities.

\subsection{Relation to stability condition for self-renewing resources}

We can also show that the same unified stability condition applies to a resource model with self-renewing resources.
Without external resource influx, resources can become extinct.
We perturb only the $M^*$ surviving resources, $R_\alpha^+\rightarrow R_\alpha^++\varepsilon\eta_\alpha^{(R)}$, and write their survival fraction as $\phi_R=M^*/M$.
As in the rest of the linear-response calculation, we assume that the perturbation is small and does not change the set of surviving resources.
Because the sum in the consumer response now contains $M^*$ rather than $M$ terms, the factor $\phi_R$ multiplies both the induced response of a surviving resource and its direct perturbation.
The consumer sensitivity is therefore

\begin{align}
\begin{split}
    \left\langle
    \left[\frac{\mathrm{d}N_0^+}{\mathrm{d}\varepsilon}\right]^2
    \right\rangle
    ={}&\frac{\phi_R}{\sigma_c^2\rho^2 \langle \chi_{R} \rangle^2}
    \left[
    \left\langle
    \left[\frac{\mathrm{d}R_0^+}{\mathrm{d}\varepsilon}\right]^2
    \right\rangle +1
    \right].
\end{split}
\label{eq:self-renewing consumer sensitivity}
\end{align}

The response of resources to perturbations in the abundances of surviving consumers remains similar to Eq.~\eqref{eq:resource half final}, albeit we again restrict it to the set of surviving resources, which gives us

\begin{align}
\begin{split}
    \left\langle
    \left[\frac{\mathrm{d}R_0^+}{\mathrm{d}\varepsilon}\right]^2
    \right\rangle
    ={}&\sigma_c^2\phi_N\gamma^{-1}
    \langle\chi_{R}^2\rangle^+
    \left[
    \left\langle
    \left[\frac{\mathrm{d}N_0^+}{\mathrm{d}\varepsilon}\right]^2
    \right\rangle+1
    \right],
\end{split}
\label{eq:self-renewing resource sensitivity}
\end{align}
where $\langle\chi_{R}^2\rangle^+$ is the analog of the second moment of the resource self-susceptibility in Eq.~\eqref{eq:stability condition}, except again restricted to the surviving resources.
Note that extinct resources have zero susceptibility.
Using this, we obtain the following relation between the first and second moments of the resource self-susceptibility, which we will use later:

\begin{align}
    \langle\chi_{R}\rangle
    =\phi_R\langle\chi_{R}\rangle^+.
\label{eq:self-renewing full susceptibility first moment}
\end{align}

\begin{align}
    \langle\chi_{R}^2\rangle
    =\phi_R\langle\chi_{R}^2\rangle^+.
\label{eq:self-renewing full susceptibility second moment}
\end{align}

Substituting Eq.~\eqref{eq:self-renewing resource sensitivity} into Eq.~\eqref{eq:self-renewing consumer sensitivity}, we get

\begin{align}
\begin{split}
    \left\langle
    \left[\frac{\mathrm{d}N_0^+}{\mathrm{d}\varepsilon}\right]^2
    \right\rangle
    ={}&
    \frac{\phi_R}{\sigma_c^2\rho^2 \langle \chi_{R} \rangle^2}
    \Biggl[
    \sigma_c^2\phi_N\gamma^{-1}
    \langle\chi_{R}^2\rangle^+
    \left[
    \left\langle
    \left[\frac{\mathrm{d}N_0^+}{\mathrm{d}\varepsilon}\right]^2
    \right\rangle+1
    \right]
    +1
    \Biggr].
\end{split}
\label{eq:self-renewing sensitivity substitution}
\end{align}

Solving for the consumer sensitivity, we get

\begin{align}
\begin{split}
    \left\langle
    \left[\frac{\mathrm{d}N_0^+}{\mathrm{d}\varepsilon}\right]^2
    \right\rangle
    =
    \frac{
    \phi_R\phi_N\gamma^{-1}\langle\chi_{R}^2\rangle^+
    +\phi_R/\sigma_c^2
    }{
    \rho^2 \langle \chi_{R} \rangle^2
    -\phi_R\phi_N\gamma^{-1}\langle\chi_{R}^2\rangle^+
    }.
\end{split}
\label{eq:self-renewing consumer sensitivity chi+}
\end{align}

Substituting our expression for $\langle\chi_{R}^2\rangle^+$ in Eq.~\eqref{eq:self-renewing full susceptibility second moment} yields

\begin{align}
\begin{split}
    \left\langle
    \left[\frac{\mathrm{d}N_0^+}{\mathrm{d}\varepsilon}\right]^2
    \right\rangle
    =
    \frac{
    \phi_N\gamma^{-1}\langle\chi_{R}^2\rangle
    +\phi_R/\sigma_c^2
    }{
    \rho^2 \langle \chi_{R} \rangle^2
    -\phi_N\gamma^{-1}\langle\chi_{R}^2\rangle
    }.
\end{split}
\label{eq:self-renewing consumer sensitivity solution}
\end{align}

the denominator in Eq.~\eqref{eq:self-renewing consumer sensitivity solution} is the same as the denominator in Eq.~\eqref{eq:alternative consumer sensitivity solution}.
This shows that point at which the response response moments diverge and the community becomes unstable is therefore still determined by the boxed unified condition \eqref{eq:stability condition}.

To simplify this further, we take the simple and typical case where all resources have the same intrinsic growth rate $b_\alpha$ and carrying capacity $K_\alpha$.
In this limit, we may assume that all surviving resources have a common self-susceptibility $\chi_{\rm common}$, while extinct resources have zero self-susceptibility.
Note that in this model, the self-susceptibilities are usually computed due to changes in the resource growth rate rather than their outflux rate.
However, this will only change the overall sign of the self-susceptibility $\chi_{\rm common}$, which will not affect the stability condition since it only involves the square of the self-susceptibility and its second moment.
Using this simplification, we can see that the $\langle \chi_{R} \rangle$ in eq. \eqref{eq:self-renewing full susceptibility first moment} can be written as

\begin{align}
    \langle \chi_{R} \rangle=\phi_R \chi_{\rm common},
\label{eq:self-renewing susceptibility first moment common}
\end{align}

while $\langle\chi_{R}^2\rangle$ from eq. \eqref{eq:self-renewing full susceptibility second moment} can be written as

\begin{align}
    \langle\chi_{R}^2\rangle=\phi_R\chi_{\rm common}^2.
\label{eq:self-renewing susceptibility second moment common}
\end{align}

Thus, this simplified limit, we see that the unified stability condition reduces to

\begin{align}
    \rho^2 > \frac{S^*}{M} \frac{\phi_R\chi_{\rm common}^2}{[\phi_R \chi_{\rm common}]^2}
    =\frac{S^*}{M} \frac{1}{\phi_R}
\label{eq:self-renewing condition almost}
\end{align}

Substituting $\phi_R = M^*/M$ yields the known stability condition for the self-renewing model~\cite{blumenthal_phase_2024}:

\begin{align}
    \rho^2 > \frac{S^*}{M*}
\label{eq:self-renewing condition}
\end{align}

Our unified stability condition further emphasizes the susceptibility of resources at or near extinction as a primary driver of stability or instability.
In the self-renewing model, the absence of external resource influx permits resource extinctions.
For extinct resources, the self-susceptibility is zero, while the surviving resources respond with susceptibility $\chi_{\rm common}$ in the limit considered above.
The first and second moments then contain different powers of the resource survival fraction, so their ratio simplifies to $M/M^*$ and gives $\rho^2>S^*/M^*$.
In the externally-supplied model, positive resource influx prevents resources from going extinct.
The resource susceptibilities therefore cannot be simply separated into surviving and extinct parts, and the second moment of the self-susceptibility must instead be evaluated over the full distribution of resources.
This is why resources with large depletion rate susceptibilities can have a disproportionate effect on the stability boundary.
The presence or absence of resource influx changes this susceptibility distribution, and hence distinguishes the stability conditions of the two resource models.

\subsection{Relation to stability condition for externally supplied resources}
To clarify the connection between the two stability conditions derived here and in the previous section, consider the expression in  Eq.~\eqref{eq:chi00 closed}. 
Squaring and taking its expectation gives us

\begin{align}
\begin{split}
    \langle \chi_{00}^2 \rangle &= 
    \left \langle \left[ -\frac{1}{2F}
    \left[
    1-\frac{f}{\sqrt{f^2-4FI_0}}
    \right] \right]^2 \right \rangle
    \\ \Rightarrow &
    4 F^2 \langle \chi_{R}^2 \rangle = 
    \left \langle \left[
    1-\frac{f}{\sqrt{f^2-4FI_0}}
    \right]^2 \right \rangle.
\end{split}
\label{eq:long bracket chi relation}
\end{align}

This equation is directly related to the long bracketed term in Eq.~\eqref{eq:big_stability_condition}. To see this, note that before evaluating the long bracketed term in Eq.~\eqref{eq:big_stability_condition} using binomial expansion, the stability condition is

\begin{align}
    \frac{1}{4\rho^2}
    \left\langle
    \left[
    1-\frac{f}{\sqrt{f^2-4FI_0}}
    \right]^2
    \right\rangle
    <\phi_N\gamma^{-1}.
\label{eq:pre-expansion stability condition}
\end{align}

Substituting eq.~\eqref{eq:long bracket chi relation} into this expression gives

\begin{align}
    \frac{F^2\langle\chi_{R}^2\rangle}{\rho^2}
    <\phi_N\gamma^{-1}.
\label{eq:pre-expansion susceptibility condition}
\end{align}

Substituting in $v_N$ from ~\eqref{eq:v_N} to $F$ in ~\eqref{eq:f_def} yields $F^2 = [\phi_N \gamma^{-1}]^2/\langle \chi_{R}^2 \rangle$. 
Finally, we can substitute this expression for $F$ into eq. \eqref{eq:pre-expansion susceptibility condition},  which gives

\begin{align}
    &
    \frac{[\phi_N \gamma^{-1}]^2 \langle \chi_{R}^2 \rangle}{\rho^2 \langle \chi_{R} \rangle^2}
    <\phi_N\gamma^{-1}
    \\ & \Rightarrow  \rho^2 > \phi_N \gamma^{-1} \frac{\langle \chi_{R}^2 \rangle}{\langle \chi_{R} \rangle^2} = \frac{S^*}{M} \frac{\langle \chi_{R}^2 \rangle}{\langle \chi_{R} \rangle^2},
\label{eq:F chi relation}
\end{align}

which is the unified stability condition in the box Eq.~\eqref{eq:stability condition}. 
The unified condition is useful when the susceptibility moment can be evaluated directly, or when we wish to compare models with different local resource dynamics.
For example, if we take the simple limit where all resources have the same susceptibility $\chi_{\text{common}}$, then the condition reduces to 

\begin{align}
    \rho^2 > \frac{S^*}{M},
\end{align}

which is comparable to the condition for self-renewing resources.

However, if we want to evaluate the susceptibilities and obtain a general stability condition for externally-supplied resources, we can plug the expression for $F$ into eq. \eqref{eq:long bracket chi relation}, rearrange the expression to make $\langle \chi_R^2 \rangle$ the subject, then plug this expression into eq. \eqref{eq:F chi relation} to yield

\begin{align}
    \rho^2 > 
    \frac{M}{4 S^*}
        \left \langle 
        \left[
        1 - \frac{b_\alpha^{\rm eff}}{
            \sqrt{(b_\alpha^{\rm eff})^2 + \frac{4 S^* I_\alpha}{M \langle \chi_R \rangle}}}
        \right]^2
        \right \rangle,
\label{eq:externally_supplied_condition_SI}
\end{align}.

This alternative derivation leaves the same average in terms of $\langle\chi_{R}^2\rangle$ rather than fully evaluating the bracket. 
If we want an explicit approximation in terms of the order parameters of the externally-supplied model, we can perform a first order binomial expansion on the bracket.
This yields the original stability condition.

\subsection{Generalisation to other forms of resource dynamics}

This alternate calculation also suggests how to treat other local resource dynamics.
For each model, one first determines the response of a single resource and then computes the first and second moments of that response over the self-consistent resource distribution.
We hypothesize that models where resources can go extinct will have a stability condition similar to the self-renewing limit, especially when resources are supplied at the same rate.
In contrast, models where resources are externally supplied and cannot go extinct will have a stability condition similar to the general unified stability condition, albeit with a different form of $\langle\chi_{R}^2\rangle$.
The hybrid model considered above provides one example that interpolates between the two limits.

More complex resource dynamics, where the influx rates depend on the current state of the community might lead to a different resource self-susceptibility distribution $\chi_{00}$ and therefore have a different stability boundary.
Cross-feeding of metabolic byproducts could be an example of such complex resource dynamics.
Further, if there were direct resource-resource interactions, we might need a slightly more general distribution over the full susceptibility matrix, rather than just the diagonal elements.

This perspective also explains what is lost when models neglect resource dynamics and replace them with effective consumer--consumer interactions.
In such descriptions, the resource susceptibility factor is usually fixed by hand, absorbed into the effective interactions, or approximated by a single value.
Thus, it is no longer amenable to feedback from consumer dynamics, and thus might lead to incorrect estimates of the stability boundary.
Keeping the resource dynamics explicit allows this factor to emerge naturally from community feedbacks, resulting in a more accurate description of	the stability boundary. 
It also cleanly decomposes how different factors such as consumer diversity, resource diversity, and resource susceptibilities impact community stability.

\section{Hybrid model with both externally-supplied and self-renewing resource supply}

In this section, we apply the unified calculation above to a model with both external resource influx and resource self-renewal. We first derive its stability condition when the influx is positive, and then show how the condition reduces to the self-renewing result when the influx is zero.

\subsection{Model set up}

\begin{align}
\begin{split}
    \frac{dR_\alpha}{dt} =&
     \underbrace{I_\alpha + R_\alpha \left[ b_\alpha - \frac{R_\alpha}{K_\alpha} \right]}_{h_\alpha (R_\alpha)} - R_\alpha \sum_{i = 1}^S c_{i \alpha} N_i.
     \\
    \frac{dN_i}{dt} =& N_i \Biggl[ \sum_{\alpha = 1}^M g_{i \alpha}  R_\alpha - d_i\Biggl].
\label{eq:hybrid model}
\end{split}
\end{align}

$I_\alpha$ is the influx rate and $b_\alpha$ is the intrinsic resource growth rate. The latter can be interpreted as the net biological growth rate minus the outflux rate, and can therefore be positive or negative.
$K_\alpha$ is the carrying capacity of resource $\alpha$ and controls the strength of resource self-inhibition.
In the calculation below, we set $K_\alpha = 1$.

\subsection{Cavity consumer and resource abundance distributions}

The equation for the cavity consumer remains unchanged from the externally-supplied model.
The cavity resource $R_0$, however, satisfies

\begin{align}
0 = 
    \mu_I + \sigma_I z_{I, 0} 
    + R_0 \left[ \mu_b - \mu_c \gamma^{-1} \langle N \rangle + \sqrt{\sigma_c^2 \gamma^{-1} \langle N^2 \rangle + \sigma_b^2} Z_R \right] 
    + R_0^2 [\sigma_g \sigma_c \rho \gamma^{-1} \langle v_{N} \rangle - 1].
    \label{eq:hybrid cavity resource dynamics}
\end{align}

Solving this equation gives the following expression for the cavity resource abundance:

\begin{align}
\begin{split}
    R_0 = & 
    - \frac{
        \mu_b - \mu_c \gamma^{-1} \langle N \rangle 
        + \sqrt{\sigma_c^2 \gamma^{-1} \langle N^2 \rangle + \sigma_b^2} Z_R}{
        2 \left[ \sigma_c \sigma_g \rho \gamma^{-1} \langle v_{N} \rangle - 1 \right]}
    \\ & - 
    \frac{\sqrt{
    [ \mu_b - \mu_c \gamma^{-1} \langle N \rangle + \sqrt{\sigma_c^2 \gamma^{-1} \langle N^2 \rangle + \sigma_b^2} Z_R ]^2 
    - 4[\sigma_c \sigma_g \rho \gamma^{-1} \langle v_{N} \rangle-1] [\mu_I + \sigma_I z_{I, 0}]
    }}{2 \left[ \sigma_c \sigma_g \rho \gamma^{-1} \langle v_{N} \rangle - 1 \right]}.
\end{split}
\label{eq:hybrid R0 distribution}
\end{align}

\subsection{Stability condition}

As above, we perturb the surviving consumers and the resources and calculate the second moments of their responses.
When $I_\alpha>0$ for every resource, resources cannot go extinct. Thus $M^*=M$ and $\phi_R=1$, and the consumer sensitivity is given by Eq.~\eqref{eq:dNde2_cavity}.
When the influx is zero, resources can go extinct. We then perturb only the $M^*$ surviving resources and use the consumer sensitivity in Eq.~\eqref{eq:self-renewing consumer sensitivity}.

We first consider $I_\alpha>0$.
For ease of notation, we write $a_0$ as

\begin{align}
    a_0
    ={}&\mu_b-\mu_c\gamma^{-1}\langle N\rangle
    +\sqrt{\sigma_c^2\gamma^{-1}\langle N^2\rangle+\sigma_b^2}\,Z_R.
\label{eq:hybrid a0 definition}
\end{align}

We will also use

\begin{align}
    H
    ={}&\sigma_g\sigma_c\rho\gamma^{-1}\langle v_{N} \rangle-1,
\label{eq:hybrid H definition}
\end{align}

and

\begin{align}
    I_0=\mu_I+\sigma_I z_{I,0}.
\label{eq:hybrid I0 definition}
\end{align}

Using this notation, Eq.~\eqref{eq:hybrid R0 distribution} becomes

\begin{align}
    R_0
    ={}&\frac{-a_0-\sqrt{a_0^2-4HI_0}}{2H}.
\label{eq:hybrid resource abundance alternative}
\end{align}

The intrinsic growth rate and the depletion generated by consumers enter $R_0$ through $a_0$ with opposite signs. An increase in $a_0$ is therefore equivalent to a decrease in the resource depletion rate. We can characterize this response using the resource self-susceptibility

\begin{align}
\begin{split}
    \chi_{R, 00}
    ={}&\frac{\partial R_0}{\partial a_0}
    =-\frac{1}{2H}
    \left[
    1+\frac{a_0}{\sqrt{a_0^2-4HI_0}}
    \right].
\end{split}
\label{eq:hybrid susceptibility}
\end{align}

Averaging over resources gives the mean susceptibility

\begin{align}
    \langle \chi_{R} \rangle=\left\langle\chi_{R, 00}\right\rangle.
\label{eq:hybrid mean susceptibility}
\end{align}

Perturbing consumer abundances changes $a_0$ through the consumer depletion term. 
Multiplying this change by $\partial R_0/\partial a_0$, squaring the result, and averaging over resources, we get

\begin{align}
\begin{split}
    \left\langle
    \left[\frac{\mathrm{d}R_0}{\mathrm{d}\varepsilon}\right]^2
    \right\rangle
    \approx{}& \sigma_c^2\phi_N\gamma^{-1}
    \left\langle\chi_{R, 00}^2\right\rangle
    \left[
    \left\langle
    \left[\frac{\mathrm{d}N_0^+}{\mathrm{d}\varepsilon}\right]^2
    \right\rangle+1
    \right]
    =
    \sigma_c^2\phi_N\gamma^{-1}
    \left\langle\chi_{R}^2\right\rangle
    \left[
    \left\langle
    \left[\frac{\mathrm{d}N_0^+}{\mathrm{d}\varepsilon}\right]^2
    \right\rangle+1
    \right].
\end{split}
\label{eq:hybrid resource sensitivity}
\end{align}

Substituting Eq.~\eqref{eq:hybrid resource sensitivity} into the consumer sensitivity in Eq.~\eqref{eq:dNde2_cavity}, and requiring the resulting second moments to remain finite, gives us the unified stability condition:

\begin{tcolorbox}[colback=white!5!white,colframe=black!100!black,
title = Hybrid model stability condition,
fonttitle = \Large\bfseries]

\begin{align}
    \rho^2>
    \frac{S^*}{M}
    \frac{\left\langle\chi_{R}^2\right\rangle}
    { \langle \chi_{R} \rangle^2}.
\label{eq:hybrid unified stability condition}
\end{align}

\end{tcolorbox}

To connect this condition to one that contains a more explicit expression for $\langle\chi_{R}^2\rangle$, we can instead square Eq.~\eqref{eq:hybrid susceptibility} and average over resources, which gives

\begin{align}
    \left\langle
    \left[
    1+\frac{a_0}{\sqrt{a_0^2-4HI_0}}
    \right]^2
    \right\rangle
    =4H^2\left\langle\chi_{R}^2\right\rangle.
\label{eq:hybrid long bracket susceptibility relation}
\end{align}

Using the same binomial expansion as in the externally-supplied calculation to evaluate these averages, Eq.~\eqref{eq:hybrid resource sensitivity} becomes

\begin{align}
\begin{split}
    \left \langle \left[ \frac{ \mathrm{d} R_0}{\mathrm{d} \varepsilon} \right]^2 \right \rangle \approx{}&
    \frac{\sigma_c^2 \phi_N \gamma^{-1}}{4 \left[ \sigma_g \sigma_c \rho \gamma^{-1} \langle v_{N} \rangle - 1 \right]^2}
     \left[ \left \langle \left[ \frac{ \mathrm{d} {N_0^+}}{\mathrm{d} \varepsilon} \right]^2 \right \rangle + 1 \right]
    \\ & \times
    \Biggl[ 1 
    + \frac{2}{\left[ \left[ \mu_b - \mu_c \gamma^{-1} \langle N \rangle \right]^2 - 4 \left[ \sigma_g \sigma_c \rho \gamma^{-1} \langle v_{N} \rangle - 1 \right] \mu_I \right]^{\frac{1}{2}}} \Biggl[ 
        \mu_b - \mu_c \gamma^{-1} \langle N \rangle
    \\ & \hphantom{\times \Biggl[ 1 - \frac{2}{\left[ \left[ \mu_b - \mu_c \gamma^{-1} \langle N \rangle \right]^2 - 4 \left[ \sigma_g \sigma_c \rho \gamma^{-1} \langle v_{N} \rangle - 1 \right] \mu_I \right]^{\frac{1}{2}}} \Biggl[} 
        - \frac{3 \left[ \mu_b - \mu_c \gamma^{-1} \langle N \rangle\right] \left[ \sigma_c^2 \gamma^{-1} \langle N^2 \rangle + \sigma_b^2 \right] }{2 \left[ \left[ \mu_b - \mu_c \gamma^{-1} \langle N \rangle \right]^2 - 4 \left[ \sigma_g \sigma_c \rho \gamma^{-1} \langle v_{N} \rangle - 1 \right] \mu_I \right]} \Biggl]
     \\ & \hphantom{\times \Biggl[}
     + \frac{1}{\left[ \mu_b - \mu_c \gamma^{-1} \langle N \rangle \right]^2 - 4 \left[ \sigma_g \sigma_c \rho \gamma^{-1} \langle v_{N} \rangle - 1 \right] \mu_I} \Biggl[ 
        \left[ \mu_b - \mu_c \gamma^{-1} \langle N \rangle \right]^2 +\sigma_c^2 \gamma^{-1} \langle N^2 \rangle + \sigma_b^2 
    \\ & \hphantom{\times \Biggl[ + \frac{1}{\left[ \mu_b - \mu_c \gamma^{-1} \langle N \rangle \right]^2 - 4 \left[ \sigma_g \sigma_c \rho \gamma^{-1} \langle v_{N} \rangle - 1 \right] \mu_I} \Biggl[}
        - \frac{5 \left[ \mu_b - \mu_c \gamma^{-1} \langle N \rangle \right]^2 \left[ \sigma_c^2 \gamma^{-1} \langle N^2 \rangle + \sigma_b^2 \right]}{\left[ \mu_b - \mu_c \gamma^{-1} \langle N \rangle \right]^2 - 4 \left[ \sigma_g \sigma_c \rho \gamma^{-1} \langle v_{N} \rangle - 1 \right] \mu_I}
    \\ & \hphantom{\times \Biggl[ + \frac{1}{\left[ \mu_b - \mu_c \gamma^{-1} \langle N \rangle \right]^2 - 4 \left[ \sigma_g \sigma_c \rho \gamma^{-1} \langle v_{N} \rangle - 1 \right] \mu_I} \Biggl[}    
    - \frac{
    3 \left[ \sigma_c^2 \gamma^{-1} \langle N^2 \rangle \right]^2
    + 6 \sigma_b^2 \left[ \sigma_c^2 \gamma^{-1} \langle N^2 \rangle \right] 
    + 3 \sigma_b^4}{\left[ \mu_b - \mu_c \gamma^{-1} \langle N \rangle \right]^2 - 4 \left[ \sigma_g \sigma_c \rho \gamma^{-1} \langle v_{N} \rangle - 1 \right] \mu_I} \Biggl]
    \Biggl].
\end{split}
\label{eq:hybrid model dR0de2}
\end{align}

Thus, Eq.~\eqref{eq:hybrid model dR0de2} is an approximate evaluation of the same second moment $\langle\chi_{R}^2\rangle$ that appears in the boxed condition. 
This condition can be evaluated using quantities that we have already determined through the self-consistency equations, whereas the boxed condition is more compact, but does not contain an explicit method to evaluate $\langle\chi_{R}^2\rangle$.

To look at this expression in various limits, we may set the resource influx to zero. 
Resources may then either go extinct or survive. 
Setting $\mu_I=\sigma_I=0$ and following similar steps as in the previous sections, we can show that 

\begin{align}
    \frac{\left\langle\chi_{R}^2\right\rangle}
    {\langle \chi_{R} \rangle^2}
    =\frac{1}{\phi_R}
    =\frac{M}{M^*}.
\label{eq:hybrid zero influx susceptibility ratio}
\end{align}

Substituting this result into Eq.~\eqref{eq:hybrid unified stability condition} gives $\rho^2>S^*/M^*$, which is the self-renewing stability condition in Eq.~\eqref{eq:self-renewing condition}. 
Thus, in the limit of no resource influx where resources can go extinct, we obtain the same stability condition in the hybrid model as in the self-renewing model. 

We may also consider two other limits in the presence of resource influx, i.e., $I_\alpha>0$, so that all resources survive and $M^*=M$.
First, disallowing resources to get diluted by setting $b_\alpha=0$ reduces Eq.~\eqref{eq:hybrid model dR0de2} to

\begin{align}
\begin{split}
    \left \langle \left[ \frac{ \mathrm{d} R_0}{\mathrm{d} \varepsilon} \right]^2 \right \rangle \approx{}&
    \frac{\sigma_c^2 \phi_N \gamma^{-1}}{4 \left[ \sigma_g \sigma_c \rho \gamma^{-1} \langle v_{N} \rangle - 1 \right]^2}
     \left[ \left \langle \left[ \frac{ \mathrm{d} {N_0^+}}{\mathrm{d} \varepsilon} \right]^2 \right \rangle + 1 \right]
    \\ & \times
    \Biggl[ 1 
    - \frac{2}{\left[ \left[ \mu_c \gamma^{-1} \langle N \rangle \right]^2 - 4 \left[ \sigma_g \sigma_c \rho \gamma^{-1} \langle v_{N} \rangle - 1 \right] \mu_I \right]^{\frac{1}{2}}} \Biggl[ 
         \mu_c \gamma^{-1} \langle N \rangle
    \\ & \hphantom{\times \Biggl[ 1 - \frac{2}{\left[ \left[  \mu_c \gamma^{-1} \langle N \rangle \right]^2 - 4 \left[ \sigma_g \sigma_c \rho \gamma^{-1} \langle v_{N} \rangle - 1 \right] \mu_I \right]^{\frac{1}{2}}} \Biggl[} 
        - \frac{3 \left[ \mu_c \gamma^{-1} \langle N \rangle\right] \left[ \sigma_c^2 \gamma^{-1} \langle N^2 \rangle \right] }{2 \left[ \left[  \mu_c \gamma^{-1} \langle N \rangle \right]^2 - 4 \left[ \sigma_g \sigma_c \rho \gamma^{-1} \langle v_{N} \rangle - 1 \right] \mu_I \right]} \Biggl]
     \\ & \hphantom{\times \Biggl[}
     + \frac{1}{\left[ \mu_c \gamma^{-1} \langle N \rangle \right]^2 - 4 \left[ \sigma_g \sigma_c \rho \gamma^{-1} \langle v_{N} \rangle - 1 \right] \mu_I} \Biggl[ 
        \left[ \mu_c \gamma^{-1} \langle N \rangle \right]^2 +\sigma_c^2 \gamma^{-1} \langle N^2 \rangle  
    \\ & \hphantom{\times \Biggl[ + \frac{1}{\left[  - \mu_c \gamma^{-1} \langle N \rangle \right]^2 - 4 \left[ \sigma_g \sigma_c \rho \gamma^{-1} \langle v_{N} \rangle - 1 \right] \mu_I} \Biggl[}
        - \frac{5 \left[ \mu_c \gamma^{-1} \langle N \rangle \right]^2 \left[ \sigma_c^2 \gamma^{-1} \langle N^2 \rangle  \right]}{\left[ \mu_c \gamma^{-1} \langle N \rangle \right]^2 - 4 \left[ \sigma_g \sigma_c \rho \gamma^{-1} \langle v_{N} \rangle - 1 \right] \mu_I}
    \\ & \hphantom{\times \Biggl[ + \frac{1}{\left[ \mu_c \gamma^{-1} \langle N \rangle \right]^2 - 4 \left[ \sigma_g \sigma_c \rho \gamma^{-1} \langle v_{N} \rangle - 1 \right] \mu_I} \Biggl[}    
    - \frac{
    3 \left[ \sigma_c^2 \gamma^{-1} \langle N^2 \rangle \right]^2}{\left[ \mu_c \gamma^{-1} \langle N \rangle \right]^2 - 4 \left[ \sigma_g \sigma_c \rho \gamma^{-1} \langle v_{N} \rangle - 1 \right] \mu_I} \Biggl]
    \Biggl].
\end{split}
\label{eq:hybrid model dR0de2 without o}
\end{align}

This has a similar structure to the stability condition for the externally supplied model in Eq.~\eqref{eq:big_stability_condition}, and shows that the presence of resource influx alone is sufficient to result in a stability condition that is rather different from the self-renewing resource model.

Second, we can also remove resource self-inhibition by setting $K_\alpha\to\infty$. 
A finite steady-state resource abundance then requires $b_\alpha<0$.
We can thus write the corresponding resource growth rate as instead a dilution rate by setting $\widetilde b_\alpha=-b_\alpha$ and denote its mean by $\mu_b$. 
With these substitutions, the resource response reduces to that of the externally-supplied model in Eq.~\eqref{eq:alternative response collected}.
Thus, removing self-inhibition while retaining positive resource influx recovers the externally-supplied model, whereas removing influx recovers the self-renewing model. 
In both cases, the resource dynamics are an integral part of the stability condition, which depends on the first and second moments of the resource self-susceptibility $\chi_{00}$.

\section{Numerical details}

All codes for running simulations and solving self-consistency equations can be found in this repository: \url{https://github.com/JamilaRowlandChandler/CRM-Resource-supply-vs-Stability}.
This repository also contains notebooks with examples for running simulations and solving self-consistency equations.

\subsection{Simulations}

The models used in simulations are of the form

\begin{align}
    \frac{\mathrm{d}R_\alpha}{\mathrm{d}t} = h_\alpha (R_\alpha) - R_\alpha \sum_{i = 1}^S c_{i \alpha} N_i + \zeta,
    \quad
    \frac{\mathrm{d}N_i}{\mathrm{d}t} =& N_i \Biggl[ \sum_{\alpha = 1}^M g_{i \alpha}  R_\alpha - d_i
    \Biggl] + \zeta,
    \label{eq:model simulations}
\end{align}

where the resource supply function $h_\alpha(R_\alpha)$ took the form 

\begin{align}
\begin{split}
    & \text{\textbf{Self-renewing : }}
    R_\alpha \left[ b_\alpha - \frac{R_\alpha}{K_\alpha} \right]
    \\
    & \text{\textbf{Externally-supplied : }}
    I_\alpha - b_\alpha R_\alpha
    \\
    & \text{\textbf{Hybrid supply : }}
    I_\alpha + R_\alpha \left[ b_\alpha - \frac{R_\alpha}{K_\alpha} \right]
\end{split}
\end{align}

Parameters were sampled from the following distributions:

\begin{align}
\begin{split}
    & c_{i \alpha} = \frac{\mu}{M} + \frac{\sigma}{\sqrt{M}} \left[ \rho z_{g, i \alpha} + \sqrt{1 - \rho^2} z_{c, i \alpha} \right], 
    \quad
    g_{i \alpha} = \frac{\mu}{M} + \frac{\sigma}{\sqrt{M}} z_{g, i \alpha},
    \quad 
    I_\alpha = I,
    \quad
    b_\alpha = 1,
    \quad
    K_\alpha = K,
    \quad
    d_i = 1,
    \\ &
    \zeta = 10^{-8},
    \label{eq:parm_sampling_simulations}
\end{split}
\end{align}

where $\zeta$ is the ``migration rate''.
This term is added to reduce numerical instability in our simulations, which occurs when some species and resource abundances are very small but not extinct.
We neglect it in our analytical calculations because it is very small.

We generated communities and simulated their dynamics in Python using \textit{numpy} and scipy's ODE solver \text{solve\char`_ivp}, and in Mathematica using \textit{NDSolve}.
For each set of parameter distributions, 20 communities were sampled.

Simulation data in fig. 2 and 3 in the main text was generated using Python.
The parameters of each community were sampled from their respective normal distributions using \textit{numpy.random.randn}.
Community dynamics were then simulated from 1 set of initial abundances sampled from $\text{Uniform}(\text{min.} = \zeta, \text{max.} = 2/M)$.
Dynamics were simulated for $t = 1000$ using the \textit{LSODA} routine from \text{scipy.integrate.solve\char`_ivp}, which is built for efficiently dealing with stiff (and non-stiff) ODE problems.
An \textit{unbounded growth} condition was also included to terminate the simulation when abundances grew unbounded.
Hyper-parameters: the solver's relative tolerance was set to $10^{-7}$, absolute tolerance to $10^{-9}$. 
Dynamics were saved at $\mathrm{d}t=5$.
Simulation data in fig. S2 was simulated in much the same way, except $t=7000$.

Simulation data in fig. S1 was generated using Mathematica.
The parameters of each community were sampled from their respective normal distributions using \text{NormalDistribution}.
Community dynamics were then simulated from 1 set of initial abundances sampled from $\text{Uniform}(\text{min.} = \zeta, \text{max.} = 2/M)$.
Dynamics were estimated using the \textit{StiffnessSwitching} routine with the \textit{ExplicitRungeKutta} method from \text{NDSolve}.
An \textit{unbounded growth} condition was also included to terminate the simulation when abundances grew unbounded.
Hyper-parameters: the solver's relative tolerance was set to $10^{-7}$, absolute tolerance to $10^{-9}$. 
Dynamics were evaluated until $t = 400$ at every $\mathrm{d}t=2$ timestep.

\textbf{Figure 2} --
For both models, 
$\rho$ varied between 0.1 and 1.0 in increments of 0.1, $\sigma$ varied between 2.0 and 12.0 in increments of 0.5.
Other shared parameters were fixed to $M = S = 150$, $\mu = 50.0$, $\mu_d=1.0$, $\sigma_d = 0.0$, $\mu_b=1.0$, $\sigma_b = 0,0$.
In the self-renewing model (Fig. 2A), $K = 1.0$.
In the externally-supplied model (Fig. 2B), $\mu_I = 1.0$, $\sigma_I = 0.0$.

\textbf{Figure 3A} -- 
From left to right, external resource influx and self-inhibition varied as follows: $\mu_I=10^{-5}$, $\sigma_I = 0.0$, $K = 1/10^{-5}$; $\mu_I=10^{-5}$, $\sigma_I = 0.0$, $K = 1$; $\mu_I=1$, $\sigma_I = 0.0$, $K = 1/10^{-5}$; $\mu_I=1$, $\sigma_I = 0.0$, $K = 1$.
Within each plot, $\rho$ varied between 0.1 and 1.0 in increments of 0.1, $\sigma$ varied between 2.0 and 12.0 in increments of 0.5.
All other parameters were fixed to $M = S = 150$, $\mu = 50.0$, $\mu_d=1.0$, $\sigma_d = 0.0$, $\mu_b=1.0$, $\sigma_b = 0.0$.

\textbf{Figure 3B} -- 
$\rho$ varied between 0.1 and 1.0 in increments of 0.1, $I_\alpha$ varied between $10^{-5}$ and $10^{0}$, where the exponent varied in increments of 0.5.
All other parameters were fixed to $M = S = 150$, $\mu = 50.0$, $\sigma = 4.0$, $\mu_d=1.0$, $\sigma_d = 0.0$, $\mu_b=1.0$, $\sigma_b = 0.0$, $K = 1$.

\textbf{Figure 3C} -- 
Simulations were run across the boundary between stability and chaos for $\rho$ and $I_\alpha$.
Values of $I_\alpha$ were the same as Fig. 3B.
For each value of $I_\alpha$, a corresponding value of $\rho$ was selected along the phase boundary.
The values of $\rho$ selected were $[0.85, 0.85, 0.8, 0.75, 0.7, 0.55, 0.4]$.
All other parameters were the same as Fig. 3B.
For each set of parameter distributions, 80 communities were sampled.

\textbf{Figure S1A}
$\rho$ varied between 0.1 and 1.0 in increments of 0.1, $\sigma$ varied between 1.0 and 2.0 in increments of 0.1.
All other parameters were fixed to $M = S = 150$, $\mu = 1.0$, $\mu_d=1.0$, $\sigma_d = 0.1$, $\mu_I=1.0$, $\sigma_I = 0.1$, $\mu_b=1.0$, $\sigma_b = 0.0$.

\textbf{Figure S2B}
$I_\alpha$ was varied between $10^{-8}$ and $10^{-0.5}$.
All other parameters were fixed to $M = S = 150$, $\rho=0.4$, $\mu = 50.0$, $\sigma=10.5$, $\mu_d=1.0$, $\sigma_d = 0.0$, $\mu_b=1.0$, $\sigma_b = 0.0$.

\subsection{Estimating community stability}

We determined the stability of each community by numerically estimating its maximum Lyapunov exponent.
In a continuous-time system, we can define the maximum Lyapunov exponent as follows \cite{strogatz_nonlinear_2018}.
Let's say we have two nearby trajectories of species and resource abundances $\x(t)$ in phase space.
The first trajectory is denoted as $\x(t)$ and the nearby trajectory is denoted as $\x(t) + \boldsymbol{\delta}(t)$, where $\boldsymbol{\delta}(0)$ is the initial distance between the trajectories.
In the limit of the initial distance $\boldsymbol{\delta}(0)$ going to zero and the time for which the two trajectories evolve going to $\infty$, we can define the maximum Lyapunov exponent $\lambda_{\max}$ in terms of the long-term distance $\boldsymbol{\delta}(t)$ between the trajectories as:

\begin{align}
\lambda_{\max} = \lim_{\substack{\delta(0) \to 0 \\ \ t \to \infty}} \frac{1}{t} \log\left[\frac{|\boldsymbol{\delta}(t)|}{|\boldsymbol{\delta}(0)|}\right]
\end{align}

We can see that if $\lambda_{\max} < 0$, the absolute distance between trajectories $|\delta(t)|$ decays over time, indicating the system converges to the same steady state.
If $\lambda_{\max} > 0$, the trajectories diverge over time, indicating the system is unstable.
We use this definition to calculate each community's maximum Lyapunov exponent using the algorithm from \cite{seydel_practical_2010}, summarised by \href{https://www.chebfun.org/examples/ode-nonlin/LyapunovExponents.html}{Hrothgar (2015)} and \href{https://github.com/Ceyron/machine-learning-and-simulation/blob/main/english/simulation_scripts/lorenz_lyapunov_exponent_numpy.ipynb}{Koehler (2024)}.

\subsubsubsection*{Algorithm for estimating $\lambda_{\max}$}

\begin{enumerate}
    \item For the CRM, extract the final species and resource abundances from the end of simulations (detailed in Appendix D.1).
    This is the ``original trajectory'' at $t=0$.
    \item Initialise a ``perturbed trajectory''.
    Perturb the original abundances by some small amount $\boldsymbol{\delta}(0)$ i.e., the Euclidean distance between the original and perturbed trajectory is $\boldsymbol{\delta}(0)$.
    \begin{itemize}
        \item We set $\boldsymbol{\delta}(0)$ to $10^{-6}$.
    \end{itemize}
    \item Simulate the dynamics of the original and perturbed trajectory, computing the normalised (log) Euclidean distance between them $\log(\boldsymbol{\delta}(t))$ at each time step. 
    Continue to simulate dynamics until $\log(\boldsymbol{\delta}(t))$ becomes approximately constant within some tolerance.
    \begin{itemize}
        \item We found simulating dynamics for $t = 1000$ was sufficient to achieve this.
    \end{itemize}
    \item To estimate the maximum Lyapunov exponent $\lambda_{\max}$, fit a line to the log distance between trajectories $\log(\boldsymbol{\delta}(t))$ in the region where it varies with time.
    The best-fit slope is the estimated maximum Lyapunov exponent $\lambda_{\max}$. 
\end{enumerate}

\subsection{Estimating community feasibility}

We observed that communities with infeasible dynamics often exhibited unbounded-like growth or were too numerically unstable for the ODE solver to handle.
These dynamics stopped the ODE solver before its set end time $t$ because it either activated the \text{unbounded\char`_growth} event function, which stops simulations when any consumer or resource grows above $10^6$, or terminated of its own accord due to numerical instability.
Therefore, we determined a community was feasible if it ran until the end time, and infeasible if it terminated at an earlier time step.

\subsection{Estimating the packing ratio}

We defined the number of surviving consumers (resources) as the number of consumers (resources) with abundances greater than extinction threshold $e$ at the end of simulations.
We arbitrarily set $e$ to $10^{-4}$.
These quantities were then used to estimate the packing ratio.

\subsection{Estimating the sensitivity of resources to changes in depletion}

To estimate the sensitivity of resources to changes in their total depletion rate $(\mathrm{d} R_\alpha / \mathrm{d} C_\alpha)^2$, as plotted in main text Fig. 3C i, we first sorted resource abundances $R_\alpha $ in order of their total depletion rates $C_\alpha = \sum_i^S c_{i, \alpha} N_i$.
We then calculated the gradient of $R_\alpha$ in terms of $C_\alpha$ using \textit{numpy.gradient} to estimate $\mathrm{d} R_\alpha / \mathrm{d} C_\alpha$.
This was then squared to obtain the $(\mathrm{d} R_\alpha / \mathrm{d} C_\alpha)^2$, then binned by their average using \textit{scipy.stats.binned.statistic}.
Hyper-parameters: the number of bins \textit{bins} was ad-hoc set to $15$.

To determine whether sensitivity of near-extinction resources to changes in their depletion affected boom-bust dynamics, we needed to estimate their maximum sensitivity.
These are plotted in Fig. 3C ii in the main text.
As shown in Fig. 3C i in the main text, resources with a total depletion rate $C_\alpha$ that is approximately equal to the intrinsic resource growth rate $b_\alpha = 1$ are near extinction.
In the chaotic phase, these resources have a spike in sensitivity to changes to their total depletion.
Therefore, we estimated the maximum sensitivity of near-extinction resources by selecting the binned sensitivities with $0.7 < C_\alpha < 1.2$ (limits were chosen \textit{ad-hoc}), found peaks in these sensitivities using \textit{scipy.signal.find\char`_peaks}, then selecting the maximum peak value.
Hyper-parameters: the threshold value at which a peak was detected \textit{threshold} ad-hoc set to $0.01$.

\subsection{Numerically solving the self-consistency equations}

The self-consistency equations describing the consumer and resource abundance distributions --- eq.s 
\eqref{eq:consumer 0th moment} - \eqref{eq:chi_R}
--- cannot be solved analytically.
Instead we needed to solve them numerically for each set of model parameter distributions.
Please see \url{https://github.com/JamilaRowlandChandler/CRM-Resource-supply-vs-Stability} to see how this numerical routine is run. 

To obtain solutions to our self-consistency equations, we used \texttt{NMinimize} using the \texttt{RandomSearch} method to run a non-linear least-squares minimisation. 
This routine traverses a large area of parameter space and does not assume the problem is convex, which helped us obtain a global solution.
This routine solved the self-consistency equations from any initial conditions within the bounds of said equations (e.g., $\langle v_{N} \rangle < 0$,  etc.), and could even solve in the infeasible region of the model.

The loss function $\mathcal{L}(\text{SCE})$ minimised was the sum of the squared difference between the estimated value of each self-consistency equation and the value calculated from the self-consistency equations listed in (\ref{box:sces}).

\begin{align}
\begin{split}
    \mathcal{L}(\text{SCE}) = \sum_{p \in \text{SCE}} (p_{i,\text{estimated}} - p_{i,\text{calculated}})^2,
    \text{ where SCE is the set } \left\{\phi_N, \langle N \rangle,  \langle N^2 \rangle,  \langle v_{N} \rangle \right\}.
\end{split}
\label{eq:loss function}
\end{align}

We then numerically evaluated $\langle R \rangle, \langle R^2 \rangle, \langle \chi_{R} \rangle$ at the best solution.

\subsubsection{Solving for the stability boundary}

Firstly, we solved the self-consistency equations using the routine we just described.
Then, we plugged the self-consistency equations into the expression describing how far the community was from the stability threshold, given by

\begin{align}
\begin{split}
&
    \frac{1}{4 \rho^2} \Biggl[
    1 
    - \frac{2}{\left[ \left[ \mu_b + \mu_c \gamma^{-1} \langle N \rangle \right]^2 - 4 \sigma_c \sigma_g \rho \gamma^{-1} \langle v_{N} \rangle \mu_I \right]^{\frac{1}{2}}} \Biggl[ 
        \mu_b + \mu_c \gamma^{-1} \langle N \rangle
    \\ & \hphantom{\frac{1}{4 \rho^2} \Biggl[ 1 - \frac{2}{\left[ \left[ \mu_b + \mu_c \gamma^{-1} \langle N \rangle \right]^2 - 4 \sigma_c \sigma_g \rho \gamma^{-1} \langle v_{N} \rangle \mu_I \right]^{\frac{1}{2}}} \Biggl[} 
        - \frac{3 \left[ \mu_b + \mu_c \gamma^{-1} \langle N \rangle\right] \left[ \sigma_c^2 \gamma^{-1} \langle N^2 \rangle + \sigma_b^2 \right] }{2 \left[ \left[ \mu_b + \mu_c \gamma^{-1} \langle N \rangle \right]^2 - 4 \sigma_c \sigma_g \rho \gamma^{-1} \langle v_{N} \rangle \mu_I \right]} \Biggl]
     \\ & \hphantom{\frac{1}{4 \rho^2} \Biggl[}
     + \frac{1}{\left[ \mu_b + \mu_c \gamma^{-1} \langle N \rangle \right]^2 - 4 \sigma_c \sigma_g \rho \gamma^{-1} \langle v_{N} \rangle \mu_I} \Biggl[ 
        \left[ \mu_b + \mu_c \gamma^{-1} \langle N \rangle \right]^2 +\sigma_c^2 \gamma^{-1} \langle N^2 \rangle + \sigma_b^2 
    \\ & \hphantom{\frac{1}{4 \rho^2} \Biggl[ + \frac{1}{\left[ \mu_b + \mu_c \gamma^{-1} \langle N \rangle \right]^2 - 4 \sigma_c \sigma_g \rho \gamma^{-1} \langle v_{N} \rangle \mu_I} \Biggl[}
        - \frac{5 \left[ \mu_b + \mu_c \gamma^{-1} \langle N \rangle \right]^2 \left[ \sigma_c^2 \gamma^{-1} \langle N^2 \rangle + \sigma_b^2 \right]}{\left[ \mu_b + \mu_c \gamma^{-1} \langle N \rangle \right]^2 - 4 \sigma_c \sigma_g \rho \gamma^{-1} \langle v_{N} \rangle \mu_I}
    \\ & \hphantom{\frac{1}{4 \rho^2} \Biggl[ + \frac{1}{\left[ \mu_b + \mu_c \gamma^{-1} \langle N \rangle \right]^2 - 4 \sigma_c \sigma_g \rho \gamma^{-1} \langle v_{N} \rangle \mu_I} \Biggl[}    
    - \frac{
    3 \left[ \sigma_c^2 \gamma^{-1} \langle N^2 \rangle
    + \sigma_b^2 \right]^2}{\left[ \mu_b + \mu_c \gamma^{-1} \langle N \rangle \right]^2 - 4 \sigma_c \sigma_g \rho \gamma^{-1} \langle v_{N} \rangle \mu_I} \Biggl]
    \Biggl] 
\\ &
    - \phi_N \gamma^{-1}.
\end{split}
\label{eq:numerical_stability_expression}
\end{align}

When this expression equals 0, the community is on the stability threshold.
For each value of $\rho$, we fitted a curve that was a function of $\sigma$ to the stability distance using \text{scipy.optimize.curve\char`_fit}.
We then interpolated the curve and found the value of $\sigma$ where the stability distance was closest to 0.

\section{Extended information}

\subsection{The self-renewing and externally-supplied model can become equivalent when resource dynamics are neglected}

Some studies have claimed that resource supply dynamics do not alter the phase transitions of the consumer-resource model ~\cite{liu_complex_2025, liu_ecosystem_2024}.
To understand why these models might be expected to exhibit equivalent phase transitions, we consider the limit where resource dynamics are fast relative to consumer-dynamics.
Resources then remain at pseudo-steady state ($\mathrm{d} R_\alpha / \mathrm{d} t = 0$), allowing their abundances to be expressed in terms of consumers and substituted into the consumer dynamics to yield a consumer-only model:

\begin{equation}
    \frac{\mathrm{d} N_i}{\mathrm{d} t} = N_i \left[ r_i - \sum_j^S A_{ij} N_j \right]
    \label{consumer-only model}
\end{equation}

\vspace{-15pt}

{
\setlength{\tabcolsep}{5pt}
\renewcommand{\arraystretch}{1.5}
\begin{table}[H]
\begin{tabular}{p{0.06\columnwidth} | p{0.42\columnwidth} p{0.35\columnwidth}}
& 
\textrm{Self-renewing} &
\textrm{Externally-supplied}
\\
\midrule
$r_i$ & 
    $\sum_\alpha^M K_\alpha b_\alpha g_{i \alpha} \Theta (R_\alpha) - d_i$ & 
    $\sum_\alpha^M [I_\alpha / b_\alpha] g_{i \alpha} - d_i$ \\
$A_{ij}$ & 
    $\sum_\alpha^M K_\alpha g_{i \alpha} c_{j \alpha} \Theta (R_\alpha)$ & 
    $\sum_\alpha^M [I_\alpha / b_\alpha^2] g_{i \alpha} c_{j \alpha}$ \\
\end{tabular}
\end{table}
}

Here, $r_i$ represents the growth rate of each consumer in the absence of others, while $A_{ij}$ represents effective pairwise interspecies interaction strengths.  For self-renewing resources, the Heaviside function $\Theta (R_\alpha)$ is 0 if resource $\alpha$ has gone extinct, and 1 otherwise.
Externally-supplied resources cannot go extinct,  consumer dynamics only reduce to eq. \eqref{consumer-only model} assuming $b_\alpha \gg \sum_i^S c_{i \alpha}N_i$.
These consumer-only models become nearly identical when $I_\alpha = b_\alpha = K_\alpha = 1$, but only in the fast-resource limit and for externally supplied resources, total consumption must be far smaller than outflux.
In general, outside of this limit, these two models could show qualitatively distinct phase transitions.

\subsection{The stability condition for externally-supplied vs self-renewing resources diverge when resource susceptibilities are not all equal}

\begin{figure}[H]
    \centering
    \includegraphics[width=0.9\textwidth]{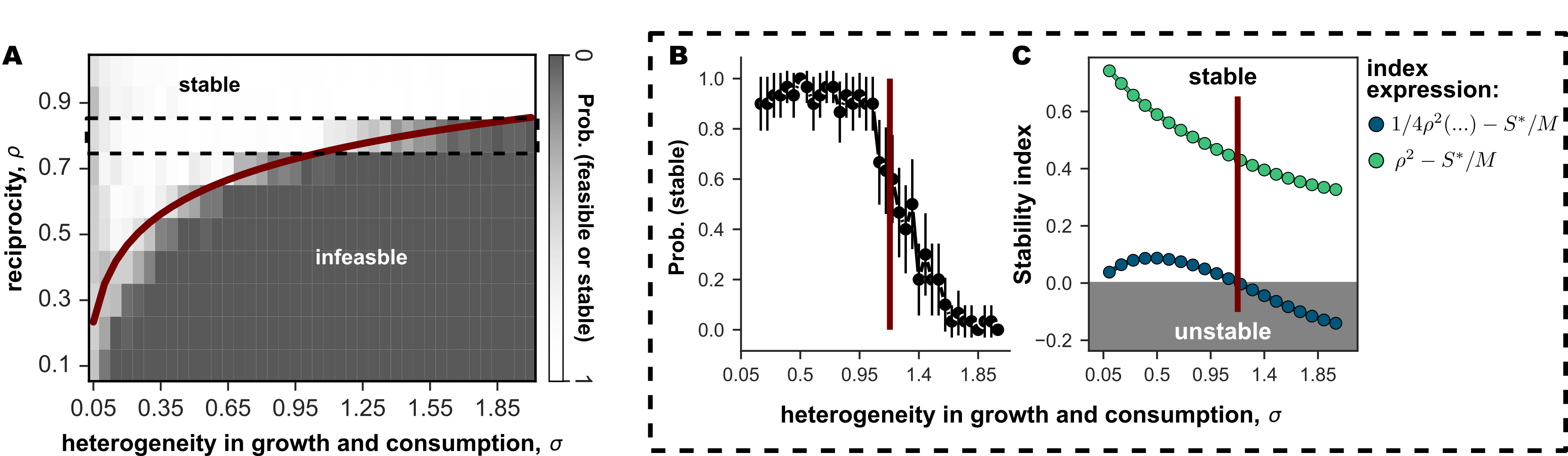}
    \caption{\textsf{
    \textbf{Ecosystems with externally-supplied resources have a different stability condition to self-renewing resources}
    (Simulation parameters are given in Appendix E.)
    \textbf{(A)}
    Phase diagram of consumer-resource interaction reciprocity ($\rho$) and heterogeneity $\sigma$ for the model with externally-supplied resources.
    Each cell shows results from 30 simulated communities: dark grey shades indicate where simulations show unstable dynamics (maximum Lyapunov exponent $> 0$) which also turn out to be infeasible in this parameter range; and white indicates the communities are stable and feasible.
    The red line is the stability threshold, analytically derived from the externally-supplied model using the cavity method.
    \textbf{(B)}
    Simulations for $\rho = 0.8$, corresponding to the outlined area in panel (A).
    Error bars represent the 95\% confidence interval, $n = 30$ communities.
    Red line = stability threshold for the externally-supplied model.
    \textbf{(C)}
    Stability indices derived under the assumption that resources have common susceptibilities (green) and with no assumptions (blue), both parametrised with the self-consistency equations solved for the externally-supplied model.
    }}
    \label{fig:S1}
\end{figure}

Here, we numerically solve the self-consistency equations for the externally-supplied model, and then estimate the stability boundary using eq. \eqref{eq:big_stability_condition}.
The ``full stability index'' for the externally-supplied model is

\begin{align}
    \dfrac{1}{4 \rho^2} \left[ 1 - \Big \langle ... \Big \rangle \right] - \dfrac{S^*}{M},
\end{align}

and is $< 0$ when the community is stable and $=0$ when the community reaches the stability threshold.
Under the simple limit where  all resources have the same susceptibilities $\chi_R$, 

\begin{align}
    \rho^2 - \dfrac{S^*}{M^*},
\end{align}

where $> 0$ when the community is stable and $=0$ when the community reaches the stability threshold.
The index in this limit is almost equivalent to the index for self-renewing resources (except resources can go extinct), but it is unclear whether the conditions remain the same or diverge when resource susceptibilities are not equal.
To ensure the susceptibilities varied across resources, we varied the resource influx rate.
Under this scenario, the full stability index captured the simulations well, but the simplified index could not: it predicted that the community should remain stable across (index $> 0$) the values of $\sigma_c$ tested, whereas the externally supplied condition correctly predicted a transition should occur (index $= 0$) (Fig. \ref{fig:S1} C).
This demonstrates that the stability condition for the externally-supplied model generally has a different form to the self-renewing condition.

\subsection{Chaos emerges in the externally-supplied model under a narrow range of resource influx rates}

\begin{figure}[H]
    \centering
    \includegraphics[width=0.9\textwidth]{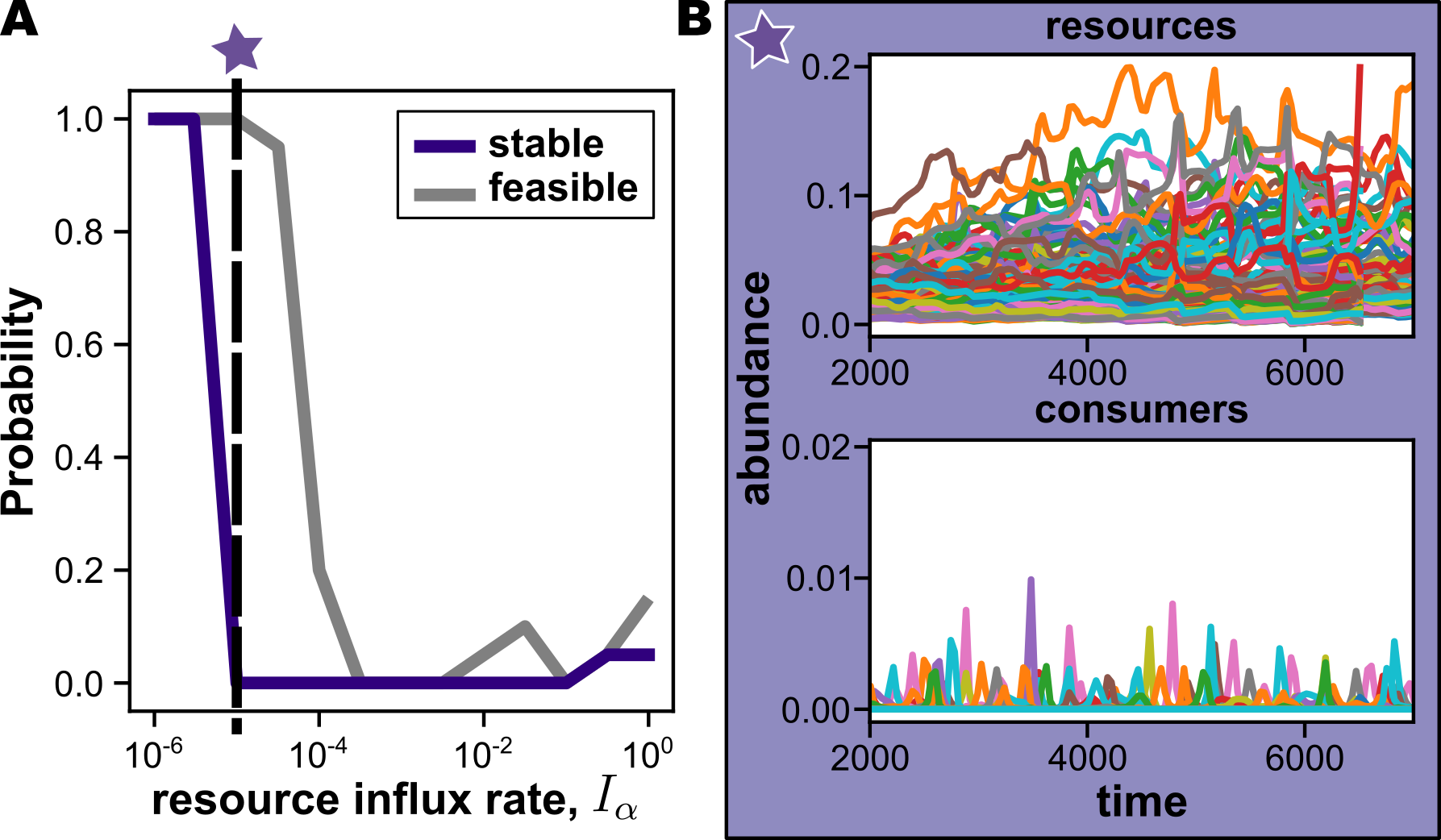}
    \caption{\textsf{
    \textbf{Chaos in the consumer-resource model with externally-supplied resources}
    (Simulation parameters are given in Appendix E.)
    \textbf{(A)}
    Phase diagram of resource influx rate $I_\alpha$ (covaried with the dilution rate $b_\alpha$) for the model with externally-supplied resources.
    Each point shows results from 20 simulated communities, showing the proportion of communities with stable (purple) or feasible (grey) dynamics.
    Purple star indicates the critical resource influx/dilution rate where unstable and feasible dynamics i.e., chaos emerges.
    \textbf{(B)}
    Representative dynamics from the chaotic region of the externally-supplied model.
    Each curve in each plot represents a single consumer or resource, as labelled.
    }}
    \label{fig:S2}
\end{figure}

\printbibliography

\end{document}